\newif\ifTechRep
\TechReptrue

\documentclass{vldb}
\usepackage{multicol}
\usepackage{graphicx}
\usepackage{balance}
\usepackage{booktabs}
\usepackage[labelformat=simple]{subcaption}
\usepackage{footnote}
\usepackage{times}
\usepackage{xspace}
\usepackage{mathtools}
\usepackage{thm-restate}
\usepackage{graphicx}
\usepackage{indentfirst}
\usepackage{booktabs}
\usepackage{dsfont}
\usepackage{footnote}
\usepackage{url}
\usepackage{dsfont}
\usepackage{bbm}
\usepackage{bm}
\usepackage{hyperref}
\usepackage{graphicx}
\usepackage{adjustbox}
\usepackage[toc]{appendix}
\usepackage[table]{xcolor}
\usepackage[labelformat=simple]{subcaption}
\usepackage{xspace}
\usepackage{graphicx}
\usepackage{pdfpages}
\usepackage{float}
\usepackage{amssymb}
\usepackage{multirow}
\usepackage[ruled,vlined]{algorithm2e}
\usepackage{enumitem}
\usepackage{soul}
\usepackage{pifont}
\usepackage{pbox}
\usepackage{cleveref}
\usepackage{makecell}
\usepackage{soul}
\DeclareFontShape{OMX}{cmex}{m}{b}{<-> cmexb10}{}
\SetSymbolFont{largesymbols}{bold}{OMX}{cmex}{m}{b}

\usepackage{amsthm, bm}
\usepackage{pdfrender}
\usepackage{soul}
\usepackage{diagbox}
\definecolor{vlightgray}{gray}{0.95}

\newcommand*{\bc}{
  \textpdfrender{
    TextRenderingMode=FillStroke,
    LineWidth=.5pt,
  }{\checkmark}
}

\SetCommentSty{mycommfont}
\setlist{nolistsep,leftmargin=*}
\theoremstyle{plain}
\theoremstyle{definition}
\newtheorem{example}{Example}[section]
\newtheorem{definition}{Definition}[section]
\newtheorem{lemma}{Lemma}[section]
\newtheorem{theorem}{Theorem}
 
\DeclareMathOperator*{\argmax}{argmax}
\SetKw{Break}{break}

\newcommand{\cm}{\checkmark}
\newcommand{\gm}{!}

\newcommand\halftab[1][.5cm]{\hspace*{#1}}
\newcommand\smallspace[1][.08cm]{\hspace*{#1}}

\newcommand{\approach}{\textsc{SQuID}\xspace}

\renewcommand{\footnotesize}{\scriptsize}
\newcommand{\queryFamily}{$\text{SPJ}_{\text{AI}}$\xspace}
\newcommand{\adb}{$\alpha$DB\xspace}
\newcommand*{\myfont}{\fontfamily{lmtt}\selectfont}

\newcommand{\sql}[1]{{\myfont{\small #1}}}
\newcommand{\myurl}[1]{{{\tiny{\footnotesize{#1}}}}}
\newcommand{\eqn}{Equation}
\newcommand{\fig}{Figure}
\newcommand{\figs}{Figures}
\newcommand{\eqns}{Equations}
\newcommand{\X}{X}
\newcommand{\x}{x}
\newcommand{\allFilterSet}{\Phi} 
\newcommand{\FilterSubset}{\varphi}
\newcommand{\prop}[1]{\langle#1\rangle}
\newcommand{\citeTechRep}{\cite{technicalReport}}
\newcommand{\theTechRep}{our technical report~\citeTechRep}

\usepackage{cleveref}
\crefformat{section}{Section~#2#1#3}
\crefformat{subsection}{Section~#2#1#3}
\crefformat{subsubsection}{Section~#2#1#3}

\hypersetup{%
  bookmarksdepth = {-2},
  pdftitle = {},
  pdfkeywords = {},
  pdfauthor = {},
  pdfstartview={FitH},
  urlcolor=black,
  linkcolor=black,
  citecolor=black,
  colorlinks=true,
}

\makeatletter
\newcommand{\textlabel}[1]{
  \phantomsection#1
  \renewcommand{\@currentlabel}{#1}
  \renewcommand{\@currentlabelname}{#1}}

\makeatother

\makeatletter
\def\thickhline{
  \noalign{\ifnum0=`}\fi\hrule \@height \thickarrayrulewidth \futurelet
   \reserved@a\@xthickhline}
\def\@xthickhline{\ifx\reserved@a\thickhline
               \vskip\doublerulesep
               \vskip-\thickarrayrulewidth
             \fi
      \ifnum0=`{\fi}}
\makeatother

\newlength{\thickarrayrulewidth}
\makeatletter
\def\@copyrightspace{\relax}
\makeatother

\makeatletter
\def\@mkbibcitation{\relax}
\makeatother

\vldbTitle{Example-Driven Query Intent Discovery: Abductive Reasoning using Semantic Similarity}
\vldbAuthors{Anna Fariha, Alexandra Meliou}
\vldbDOI{https://doi.org/10.14778/xxxxxxx.xxxxxxx}
\vldbVolume{12}
\vldbNumber{xxx}
\vldbYear{2019}

\begin{document}

\title{Example-Driven Query Intent Discovery:\\
Abductive Reasoning using Semantic Similarity}
\numberofauthors{2} 
\author{
\alignauthor
Anna Fariha\\
       \affaddr{College of Information and Computer Sciences}\\
       \affaddr{University of Massachusetts Amherst}\\
       \email{afariha@cs.umass.edu}
\alignauthor
 Alexandra Meliou\\
       \affaddr{College of Information and Computer Sciences}\\
       \affaddr{University of Massachusetts Amherst}\\
       \email{ameli@cs.umass.edu}
}

\maketitle

\begin{abstract}

\looseness-1
Traditional relational data interfaces require precise structured queries over
potentially complex schemas. These rigid data retrieval mechanisms pose hurdles
for non-expert users, who typically lack language expertise and are unfamiliar
with the details of the schema. \emph{Query by Example} (QBE) methods offer an
alternative mechanism: users provide examples of their intended query output
and the QBE system needs to infer the intended query. However, these approaches
focus on the structural similarity of the examples and ignore the richer
context present in the data. As a result, they typically produce queries that
are too general, and fail to capture the user's intent effectively. In this
paper, we present \approach, a system that performs \emph{semantic
similarity-aware query intent discovery}. Our work makes the following
contributions:
(1)~We design an end-to-end system that automatically formulates
select-project-join queries in an \emph{open-world} setting, with optional
group-by aggregation and intersection operators; a much larger class than prior
QBE techniques.
(2)~We express the problem of query intent discovery using a
\emph{probabilistic abduction model}, that infers a query as the most likely
explanation of the provided examples.
(3)~We introduce the notion of an \emph{abduction-ready} database, which
precomputes semantic properties and related statistics, allowing \approach to
achieve real-time performance.
(4)~We present an extensive empirical evaluation on three real-world datasets,
including user-intent case studies, 
demonstrating that \approach is
efficient and effective, and outperforms machine learning methods, as well as the state-of-the-art in the related
query reverse engineering problem.

\end{abstract}

\section{Introduction}

\looseness-1 Database technology has expanded drastically, and its audience has
broadened, bringing on a new set of usability requirements. A significant group
of current database users are non-experts, such as data enthusiasts and
occasional users. These non-expert users want to explore data, but lack the
expertise needed to do so. Traditional database technology was not designed
with this group of users in mind, and hence poses hurdles to these non-expert
users. Traditional query interfaces allow data retrieval through
well-structured queries. To write such queries, one needs expertise in the
query language (typically SQL) and knowledge of the, potentially complex,
database schema. Unfortunately, occasional users typically lack both. Query by
Example (QBE) offers an alternative retrieval mechanism, where users specify
their intent by providing example tuples for their query
output~\cite{MottinVLDB2016}.

Unfortunately, traditional QBE systems~\cite{ShenSIGMOD2014,
PsallidasSIGMOD2015, DeutchICDE2016} for relational databases make a strong and
oversimplifying assumption in modeling user intent: they implicitly treat the
structural similarity and data content of the example tuples as the only
factors specifying query intent. As a result, they consider all queries that
contain the provided example tuples in their result set as equally likely to
represent the desired intent.\footnote{More nuanced QBE systems exist, but
typically place additional requirements or significant restrictions over the
supported queries (Figure~\ref{relatedWorkMatrix}).} This ignores the richer
context in the data that can help identify the intended query more effectively.

\begin{figure}[t]   
    \begin{subtable}[b]{0.21\textwidth}
    \centering
    \renewcommand{\arraystretch}{0.9}
   { \small
        \begin{tabular}{|l|l|}
            \multicolumn{2}{c}{\sql{academics}}\\
            \hline
            \rowcolor{vlightgray}
            \multicolumn{1}{|c|}{id} & 
            \multicolumn{1}{c|}{name}\\
            \hline
                100 & Thomas Cormen\\
                \textbf{101} & \textbf{Dan Suciu}\\
                102 & Jiawei Han\\
                \textbf{103} & \textbf{Sam Madden}\\
                104 & James Kurose\\
                105 & Joseph Hellerstein \\
            \hline
        \end{tabular}}
    \end{subtable}
    \hspace{2mm}
    \begin{subtable}[b]{0.24\textwidth}
    \centering
    \renewcommand{\arraystretch}{0.9}
    {\small
        \begin{tabular}{|l|l|}
            \multicolumn{2}{c}{\sql{research}}\\
            \hline
            \rowcolor{vlightgray}
            \multicolumn{1}{|c|}{aid} & 
            \multicolumn{1}{c|}{interest}\\
            \hline
                100 & algorithms\\
                \textbf{101} & \underline{data management}\\
                102 & data mining\\             
                \textbf{103} & \underline{data management}\\
                103 & distributed systems\\
                104 & computer networks\\
                105 & data management\\
                105 & distributed systems\\
            \hline
        \end{tabular}}
    \end{subtable}
    \vspace{-2mm}
    \caption{Excerpt of two relations of the CS Academics database.  
    Dan Suciu and Sam Madden (in bold), both have 
    research interests in data management.}
    \label{collegeDB}
    \vspace{-5mm}
\end{figure}

\begin{example}\label{collegeDBExample}
In \fig~\ref{collegeDB}, the relations \sql{academics} and \sql{research} store information
about CS researchers and their research
interests. Given the user-provided set of examples \sql{\{Dan Suciu, Sam
Madden\}}, a human can posit that the user is likely looking for \sql{data
management} researchers. However, a QBE system, that looks for queries based only
on the structural similarity of the examples, produces \sql{Q1} to capture
the query intent, which is too general:\\ 
\small
\indent \sql{Q1:} \sql{SELECT name FROM academics}\\
\normalsize
\noindent In fact, the QBE system will generate the same generic query
\sql{Q1} for any set of names from the relation \sql{academics}. Even
though the intended semantic context is present in the data (by associating
academics with research interest information using the relation
\sql{research}), existing QBE systems fail to capture it. The more specific
query that better represents the semantic similarity among the example tuples
is \sql{Q2}:\\
\small
\indent \sql{Q2:} \sql{SELECT name FROM academics, research\\
\indent \halftab \smallspace  \smallspace  WHERE research.aid = academics.id AND\\
\indent \halftab \halftab  \halftab \smallspace \smallspace   research.interest = `data management'}
\normalsize
\end{example}

\noindent Example~\ref{collegeDBExample} shows how reasoning about the semantic
similarity of the example tuples can guide the discovery of the correct
query structure (join of the \sql{academics} and \sql{research} tables),
as well as the discovery of the likely intent (research interest in
\sql{data management}).

We can often capture semantic similarity through direct attributes of the
example tuples. These are attributes associated with a tuple within the same
relation, or through simple key-foreign key joins (such as research interest in
Example~\ref{collegeDBExample}). Direct attributes capture intent that is
\emph{explicit}, precisely specified by the particular attribute values.
However, sometimes query intent is more vague, and not expressible by explicit
semantic similarity alone. In such cases, the semantic similarity of the
example tuples is \emph{implicit}, captured through deeper associations with
other entities in the data (e.g., type and quantity of movies an actor appears
in).

\begin{figure}[t]
    \begin{center}
    \includegraphics[trim={0mm 0mm 0mm -4mm}, width=240px]{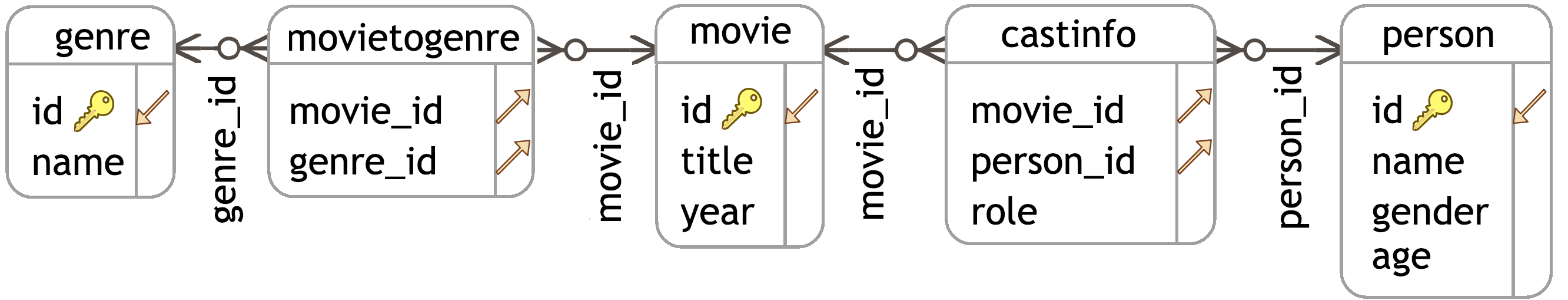}
    \vspace{-5mm}
    \caption{Partial schema of the IMDb database. The schema contains 2 entity
    relations: \sql{movie}, \sql{person}; and a semantic property
    relation: \sql{genre}. The relations \sql{castinfo} and
    \sql{movietogenre} associate entities and semantic properties.}
    \label{schema}
    \end{center}
    \vspace{-6mm}
\end{figure}

\begin{example}\label{multipleIntent} 
    
The IMDb dataset contains a wealth of information related to the movies and
entertainment industry. We query the IMDb dataset (\fig~\ref{schema}) with a
QBE system, using two different sets of examples:

\noindent
{\small
\begin{tabular}{lcl}
    		 \sql{ET1=}\{\sql{Arnold Schwarzenegger} &\phantom{xx} & \sql{ET2=}\{\sql{Eddie Murphy}\\
    \phantom{\sql{ET1=}\{}\sql{Sylvester Stallone} && \phantom{\sql{ET1=}\{}\sql{Jim Carrey}\\
    \phantom{\sql{ET1=}\{}\sql{Dwayne Johnson}\} && \phantom{\sql{ET1=}\{}\sql{Robin Williams}\}
\end{tabular}
}

\noindent \sql{ET1} contains the names of three actors from a public list of
``physically strong'' actors\footnote{\url{https://www.imdb.com/list/ls050159844}};
\sql{ET2} contains the names of three actors from a public list of ``funny''
actors\footnote{\url{https://www.imdb.com/list/ls000025701}}. \sql{ET1} and \sql{ET2} represent different query
intents (\emph{strong} actors and \emph{funny} actors, respectively), but a
standard QBE system produces the same generic query for both:\\
\small
\indent \sql{Q3:} \sql{SELECT person.name FROM person}
\normalsize

\noindent
Explicit semantic similarity cannot capture these different intents, as
there is no attribute that explicitly characterizes an actor as ``strong'' or
``funny''. Nevertheless, the database encodes these associations implicitly,
in the number and type of movies an actor appears in (``strong'' actors
frequently appear in action movies, and ``funny'' actors in comedies).
\end{example}

\noindent Standard QBE systems typically produce queries that are too general,
and fail to capture nuanced query intents, such as the ones in
Examples~\ref{collegeDBExample} and~\ref{multipleIntent}. Some prior approaches
attempt to refine the queries based on additional, external information, such
as external ontologies~\cite{LimEDBT2013}, provenance information of the
example tuples~\cite{DeutchICDE2016}, and user feedback on multiple (typically
a large number) system-generated examples~\cite{BonifatiTODS2016, LiVLDB2015,
DimitriadouTKDE2016}. Other work relies on a \emph{closed-world}
assumption\footnote{In the closed-world setting, a tuple not specified as an
example output is assumed to be excluded from the query result.} to produce
more expressive queries~\cite{LiVLDB2015, WangPLDI2017, ZhangASE2013} and thus
requires complete examples of input databases and output results. Providing
such external information is typically complex and tedious for a non-expert.

\looseness-1 In contrast with prior approaches, in this paper, we propose a
method and present an end-to-end system for discovering query intent
effectively and efficiently, in an \emph{open-world} setting, \emph{without the
need for any additional external information}, beyond the initial set of
example tuples.\footnote{Figure~\ref{relatedWorkMatrix} provides a summary
exposition of prior work, and contrasts with our contributions. We detail this
classification and metrics in \ifTechRep Appendix~\ref{app:matrix}\else
\theTechRep\fi\ and discuss the related work in Section~\ref{relatedWork}.}
\approach, our semantic similarity-aware query intent discovery
framework~\cite{DBLP:conf/sigmod/FarihaSM18}, relies on two key insights:
(1)~It exploits the information and associations already present in the data to
derive the explicit and implicit similarities among the provided examples.
(2)~It identifies the significant semantic similarities among them using
\emph{abductive reasoning}, a logical inference mechanism that aims to derive a
query as the simplest and most likely explanation for the observed results
(example tuples). We explain how \approach uses these insights to handle the
challenging scenario of Example~\ref{multipleIntent}.

\newcommand{\eln}{12}
\newcommand{\sln}{3}

\begin{figure}[t]       
    \setlength\tabcolsep{1pt}
    \centering
    \resizebox{\columnwidth}{!}{
        \small
        \begin{tabular}{|c|c|p{30mm}|c|c|c|c|c|c|c|c|l|}
        \hline
                             \multicolumn{3}{|c|}{\cellcolor{vlightgray}{\multirow[c]{2}{*}{\cellcolor{vlightgray}{\rotatebox[origin=lb]{0}{{}}}}}} & \multicolumn{4}{c|}{{\multirow[c]{2}{*}{\rotatebox[origin=lb]{0}{{query class}}} }} & &&&&\\
							 \multicolumn{3}{|c|}{\cellcolor{vlightgray}\raisebox{1.6ex}[1.6ex]{\textbf{Legend}}} & \multicolumn{4}{c|}{} & & & & &\\
		\cline{4-7}
                             \multicolumn{2}{|r}{\cellcolor{vlightgray}{\small QBE:}} & \multicolumn{1}{l|}{\cellcolor{vlightgray}\small Query by Example} & &&& & &&&&\\

        \multicolumn{2}{|r}{\cellcolor{vlightgray}{\small QRE:}} & \multicolumn{1}{l|}{\cellcolor{vlightgray}\small Query Reverse Engineering} &&&&&&&&& \multicolumn{1}{c|}{additional}\\
        \multicolumn{2}{|r}{\cellcolor{vlightgray}{\small DX:}} & \multicolumn{1}{l|}{\cellcolor{vlightgray}\small Data Exploration} &&&&&&&&& \multicolumn{1}{c|}{requirements}\\
        \multicolumn{2}{|r}{\cellcolor{vlightgray}{\small KG:}} & \multicolumn{1}{l|}{\cellcolor{vlightgray}\small Knowledge Graph} &&&&&&&&&\\
        \multicolumn{2}{|r}{\cellcolor{vlightgray}{\gm:}} & \multicolumn{1}{l|}{\cellcolor{vlightgray}\small with significant restrictions}
                            & 
                            \multirow[t]{4}{*}{\rotatebox[origin=lb]{90}{{\small join}}} &
                            \multirow[t]{4}{*}{\rotatebox[origin=lb]{90}{{\small projection}}} &
                            \multirow[t]{4}{*}{\rotatebox[origin=lb]{90}{{\small selection}}} &
                            \multirow[t]{4}{*}{\rotatebox[origin=lb]{90}{{\small aggregation}}} &
                            \multirow[t]{6}{*}{\rotatebox[origin=lb]{90}{{\small semi-join}}} &
                            \multirow[t]{6}{*}{\rotatebox[origin=lb]{90}{{\small implicit property}}} &
                            \multirow[t]{6}{*}{\rotatebox[origin=lb]{90}{{\small scalable}}} &
                            \multirow[t]{6}{*}{\rotatebox[origin=lb]{90}{{\small open-world}}} &\\

        \hline
        \hline
                                 
        \multirow{9}{*}{\rotatebox[origin=lb]{90}{QBE}}
        & \multirow{5}{*}{\rotatebox[origin=lb]{90}{{\small relational}}}  
                             & \textbf{\approach}                          & \bc   & \bc  & \bc  & \bc    & \bc & \bc     & \bc & \bc &                           \\ \cline{\sln-\eln}                                                                                            
                             && Bonifati et al.~\cite{BonifatiTODS2016}      & \cm   & \cm  & !     &      &  \cm   &       & \cm & \cm &   {\small user feedback}                \\ \cline{\sln-\eln} 
                             && QPlain~\cite{DeutchICDE2016}                 & \cm   & \cm  & \cm  &      & \cm &       & \cm & \cm &   {\small provenance input}              \\ \cline{\sln-\eln} 
                             && Shen et al.~\cite{ShenSIGMOD2014}            & \cm   & \cm  &      &      &     &       & \cm & \cm &                           \\ \cline{\sln-\eln} 
                             && \textsc{FastTopK}~\cite{PsallidasSIGMOD2015} & \cm   & \cm  &      &      &     &       & \cm & \cm &                           \\ \cline{2-\eln}                                                                                                                                                                                                                                           
        & \multirow{4}{*}{\rotatebox[origin=lb]{90}{{\small KG}}}                                                             
                             & Arenas et al.~\cite{SPARQLRE}                & \cm   & \cm  & \gm  &      & \cm &        & \cm & \cm &                           \\ \cline{\sln-\eln} 
                             && SPARQLByE~\cite{SPARQLByE2016}               & \cm   & \cm  & \gm  &      & \cm &       & \cm & \cm &    {\small negative examples}  \\ \cline{\sln-\eln} 
                             && \textsc{GQBE}~\cite{JayaramTKDE2015}         & \cm   & \cm  & \gm  &      & \cm &       & \cm & \cm &                           \\ \cline{\sln-\eln} 
                             && QBEES~\cite{MetzgerJIIS2017}                 & \cm   & \cm  & \gm  &      & \cm &       & \cm & \cm &                           \\ \hline\hline
        \multirow{9}{*}{\rotatebox[origin=lb]{90}{QRE}}  
        & \multirow{9}{*}{\rotatebox[origin=lb]{90}{{\small relational}}}                                                                                     
                             & PALEO-J~\cite{PanevEDBT2016}                 & \cm   & \cm  & \cm  & \cm  &     &     & \cm &     &   {\small top-k queries only}             \\ \cline{\sln-\eln} 
                             && SQLSynthesizer~\cite{ZhangASE2013}           & \cm   & \cm  & \cm  & \cm  & \cm &       &     &     &   {\small schema knowledge}        \\ \cline{\sln-\eln} 
                             && \textsc{Scythe}~\cite{WangPLDI2017}          & \cm   & \cm  & \cm  & \cm  & \cm &       &     &     &   {\small schema knowledge}        \\ \cline{\sln-\eln} 
                             && Zhang et al.~\cite{ZhangSIGMOD2013}          & \cm   &      &      &      &     &       & \cm & \cm &                           \\ \cline{\sln-\eln} 
                             && \textsc{REGAL}~\cite{TanVLDB2017}            &       & \cm  & \cm  & \cm  &     &    & \cm &     &                          \\ \cline{\sln-\eln} 
                             && \textsc{REGAL+}~\cite{regalPlus2018}         & \cm   & \cm  & \cm  & \cm  &     &       &     &     &                           \\ \cline{\sln-\eln} 
                             && \textsc{FastQRE}~\cite{FastQRE}              & \cm   & \cm  &      &      &     &       & \cm & \cm &                           \\ \cline{\sln-\eln} 
                             && QFE~\cite{LiVLDB2015}                        & \cm   & \cm  & \cm  &      &     &       &     &     &   {\small user feedback}            \\ \cline{\sln-\eln} 
                             && \textsc{TALOS}~\cite{TranVLDB2014}           & \cm   & \cm  & \cm  & \cm  & \cm &       & \cm &     &                           \\ \hline\hline
        
        \multirow{2}{*}{\rotatebox[origin=lb]{90}{DX}}  
        & \multirow{2}{*}{\rotatebox[origin=lb]{90}{{\small rel.}}}                                                            
                             & AIDE~\cite{DimitriadouTKDE2016}              &       &     & \cm  &      &     &      & \cm & \cm &    {\small user feedback}              \\ \cline{\sln-\eln} 
                             && REQUEST~\cite{GeBigdata2016}                 &       &      & \cm  &      &     &    & \cm & \cm &    {\small user feedback}              \\ \hline 
                 
        \end{tabular}
    }
    \vspace{-2mm}
    \caption{\approach captures complex intents and more expressive queries than prior work in the open-world setting.}
    \vspace{-4mm}
    \label{relatedWorkMatrix}
\end{figure}

\begin{example}\label{funnyActorExample} We query the IMDb dataset with
\approach, using the example tuples in \sql{ET2}
(Example~\ref{multipleIntent}). \approach discovers the following semantic
similarities among the examples: (1) all are \sql{Male}, (2) all are
\sql{American}, and (3) all appeared in more than 40 \sql{Comedy} movies. Out
of these properties, \sql{Male} and \sql{American} are very common in the IMDb
database. In contrast, a very small fraction of persons in the dataset are
associated with such a high number of \sql{Comedy} movies; this means that it
is unlikely for this similarity to be coincidental, as opposed to the other
two. Based on abductive reasoning, \approach selects the third semantic
similarity as the best explanation of the observed example tuples, and produces
the query:\\
\small
\noindent \smallspace \smallspace \sql{Q4:} \noindent \sql{SELECT person.name\\
\noindent \smallspace \smallspace \halftab \smallspace \smallspace FROM person, castinfo, movietogenre, genre\\
\noindent \smallspace \smallspace \halftab \smallspace \smallspace WHERE person.id = castinfo.person\_id\\
\noindent \smallspace \smallspace \halftab \halftab AND castinfo.movie\_id = movietogenre.movie\_id\\
\noindent \smallspace \smallspace \halftab \halftab AND movietogenre.genre\_id = genre.id\\
\noindent \smallspace \smallspace \halftab \halftab AND genre.name = `Comedy'\\
\noindent \smallspace \smallspace \halftab \smallspace \smallspace GROUP BY person.id\\
\noindent \smallspace \smallspace \halftab \smallspace \smallspace HAVING count(*) >= 40}
\normalsize
\end{example}

In this paper, we make the following contributions:
\begin{itemize}
      
	  \item We design an end-to-end system, \approach, that automatically
	  formulates select-project-join queries with optional group-by aggregation
	  and intersection operators (\queryFamily) based on few user-provided example
	  tuples. \approach does not require the users to have any knowledge of the
	  database schema or the query language. In contrast with existing approaches,
	  \approach does not need any additional user-provided information, and
	  achieves very high precision with very few examples in most cases.

      \item \approach infers the \emph{semantic similarity} of the example
      tuples, and models query intent using a collection of \emph{basic} and
      \emph{derived} semantic property \emph{filters}
      (\cref{semanticPropertyFilterSection}). Some prior work has explored the
      use of semantic similarity in knowledge graph retrieval
      tasks~\cite{ZhangSIGIR2017, MetzgerJIIS2017, JayaramTKDE2015}. However,
      these prior systems do not directly apply to the relational domain, and
      do not model implicit semantic similarities, derived from aggregating
      properties of affiliated entities (e.g., number of comedy movies an actor
      appeared in).
      
      \item We express the problem of query intent discovery using a
      \emph{probabilistic abduction model} (\cref{model}). This model allows
      \approach to identify the semantic property filters that represent the
      most probable intent given the examples.
      
      \item \approach achieves real-time performance through an offline
      strategy that pre-computes semantic properties and related statistics to
      construct an \emph{abduction-ready} database
      (\cref{offlinePrecomputation}). During the online phase, \approach
      consults the abduction-ready database to derive relevant semantic
      property filters, based on the provided examples, and applies abduction
      to select the optimal set of filters towards query intent discovery
      (\cref{intentDiscovery}). We prove the correctness of the abduction
      algorithm in Theorem~\ref{theTheorem}.

      \item Our empirical evaluation includes three real-world datasets, 41
      queries covering a broad range of complex intents and structures, and
      three case studies (\cref{experiment}). We further compare with
      TALOS~\cite{TranVLDB2014}, a state-of-the-art system that captures very
      expressive queries, but in a closed-world setting. We show that \approach
      is more accurate at capturing intent and produces better queries, often
      reducing the number of predicates by orders of magnitude. We also
      empirically show that \approach outperforms a semi-supervised machine
      learning system~\cite{elkan2008learning}, which learns classification
      models from positive examples and unlabeled data.
\end{itemize}

\section{{\large\textbf{\approach}} overview}\label{background}

In this section, we first discuss the challenges in example-driven query
intent discovery and highlight the shortcomings of existing approaches. We
then formalize the problem of query intent discovery using a probabilistic
model and describe how \approach infers the most likely query intent using
abductive reasoning. Finally, we present the system architecture for
\approach, and provide an overview of our approach.

\subsection{The Query Intent Discovery Problem}\label{sec:QuIDprob}

\approach aims to address three major challenges that hinder existing QBE systems:

\textbf{Large search space.} Identifying the intended query given
a set of example tuples can involve a huge search space of potential candidate
queries. Aside from enumerating the candidate queries, validating them is
expensive, as it requires executing the queries over potentially very large
data. Existing approaches limit their search space in three ways: 
(1)~They often focus on project-join (PJ) queries only. Unfortunately, ignoring
selections severely limits the applicability and practical impact of these
solutions.
(2)~They assume that the user provides a large number of examples or
interactions, which is often unreasonable in practice.
\begin{sloppy}
(3)~They make a closed-world assumption, thus needing complete sets of input
data and output results. In contrast, \approach focuses on a much larger and
more expressive class of queries, \emph{select-project-join queries with
optional group-by aggregation and intersection operators}
(\queryFamily)\footnote{The \queryFamily queries derived by \approach limit
joins to key-foreign key joins, and conjunctive selection predicates of the
form \sql{attribute OP value}, where \sql{OP} $\in$ $\{=, \ge, \le\}$ and
\sql{value} is a constant.}, and is effective in the open-world setting with
very few examples.
\end{sloppy}

\textbf{Distinguishing candidate queries.}
In most cases, a set of example tuples does not uniquely identify the target
query, i.e., there are multiple valid queries that contain the example tuples
in their results. Most existing QBE systems do not distinguish among the valid
queries~\cite{ShenSIGMOD2014} or only rank them according to the degree of
input containment, when the example tuples are not fully contained by the
query output~\cite{PsallidasSIGMOD2015}. In contrast, \approach exploits the
semantic context of the example tuples and ranks the valid queries based on a
probabilistic abduction model of query intent.

\textbf{Complex intent.}
A user's information need is often more complex than what is explicitly
encoded in the database schema (e.g., Example~\ref{multipleIntent}). Existing
QBE solutions focus on the query structure and are thus ill-equipped to
capture nuanced intent. While \approach still produces a structured query in
the end, its objectives focus on capturing the semantic similarity of the
examples, both explicit and implicit. \approach thus draws a contrast between
the traditional query-by-example problem, where the query is assumed to be the
hidden mechanism behind the provided examples, and the \emph{query intent
discovery problem} that we focus on in this work.

 \begin{figure}[t] 
     \begin{center}
         \includegraphics[width=\columnwidth]{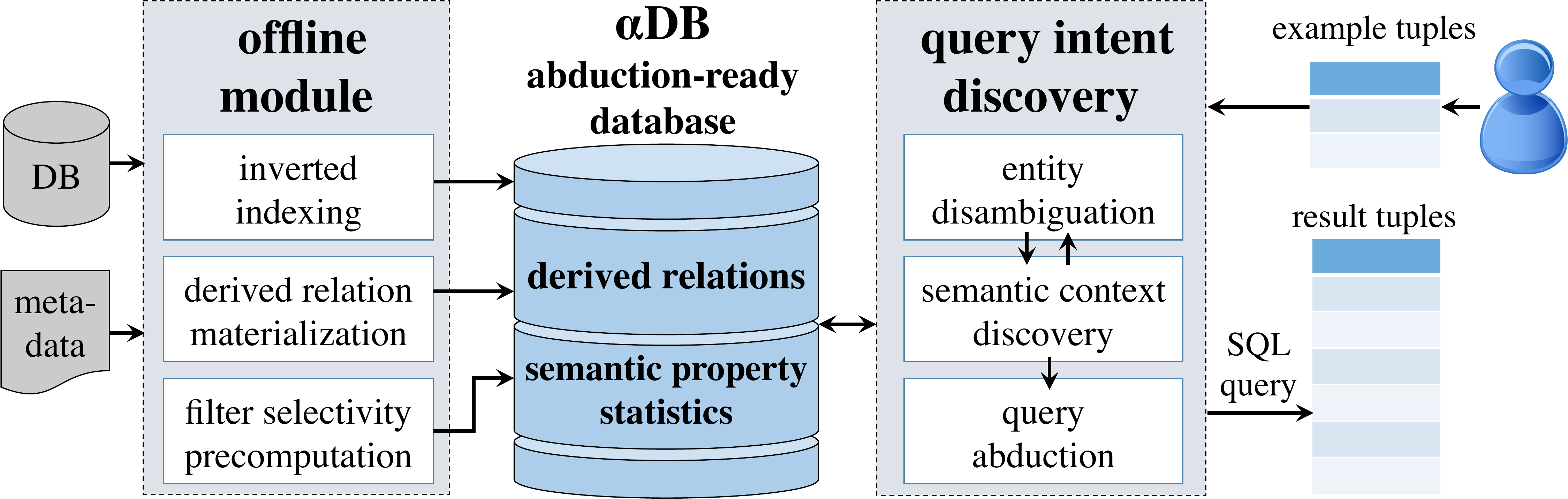}
	     \vspace{-5mm}
         \caption{
         \approach's operation includes an offline module, which constructs an
         \emph{abduction-ready} database (\adb) and precomputes statistics of
         semantic properties. During normal operation, \approach's query intent
         discovery module interacts with the \adb to identify the semantic
         context of the provided examples and abduces the most likely query
         intent.}
         \label{systemArchitecture} 
     \end{center}
     \vspace{-5mm} 
 \end{figure} 
     
\smallskip

We proceed to formalize the problem of query intent discovery. We use
$\mathcal{D}$ to denote a database, and $Q(\mathcal{D})$ to denote the set of
tuples in the result of query $Q$ operating on $\mathcal{D}$.

\begin{definition}[Query Intent Discovery]\label{def:QuID}
    For a database
$\mathcal{D}$ and a user-provided example tuple set $E$, the query intent
discovery problem is to find an {\queryFamily} query $Q$ such that:\\
\indent $\bullet$ $E \subseteq Q(\mathcal{D})$\\
\indent $\bullet$ $Q = \argmax_q Pr(q|E)$
\end{definition}

More informally, we aim to discover an {\queryFamily} query $Q$ that contains $E$ within its
result set and maximizes the query posterior, i.e., the conditional probability
$Pr(Q|E)$.

\subsection{Abductive Reasoning}\label{abductiveReasoning} 

\approach solves the query intent discovery problem (Definition~\ref{def:QuID})
using \emph{abduction}. Abduction or abductive
reasoning~\cite{abductionMenzies96, DBLP:reference/ml/Kakas17, BertossiS17,
ArieliDNB04} refers to the method of inference that finds the best explanation
(query intent) of an often incomplete observation (example tuples). Unlike
deduction, in abduction, the premises do not guarantee the conclusion. So, a
deductive approach would produce all possible queries that contain the example
tuples in their results, and it would guarantee that the intended query is one
of them. However, the set of valid queries is typically extremely large,
growing exponentially with the number of properties and the size of the data
domain. In our work, we model query intent discovery as an abduction problem
and apply abductive inference to discover the most likely query intent. More
formally, given two possible candidate queries, $Q$ and $Q^\prime$, we infer
$Q$ as the intended query if $Pr(Q|E) > Pr(Q^\prime|E)$.

\begin{example}\label{ex11} Consider again the scenario of
Example~\ref{collegeDBExample}. \approach identifies that the two example
tuples share the semantic context \sql{interest = data management}. \sql{Q1}
and \sql{Q2} both contain the example tuples in their result set. However, the
probability that two tuples drawn randomly from the output of \sql{Q1} would
display the identified semantic context is low ($(\frac{3}{7})^2\approx 0.18$
in the data excerpt). In contrast, the probability that two tuples drawn
randomly from the output of \sql{Q2} would display the identified semantic
context is high ($1.0$). Assuming that \sql{Q1} and \sql{Q2} have equal priors
($Pr(\sql{Q1}) = Pr(\sql{Q2})$), then from Bayes' rule $Pr(\sql{Q2}|E) >
Pr(\sql{Q1}|E)$. \end{example}

\begin{figure}[t]
    \centering
    \includegraphics[width=0.48\textwidth]{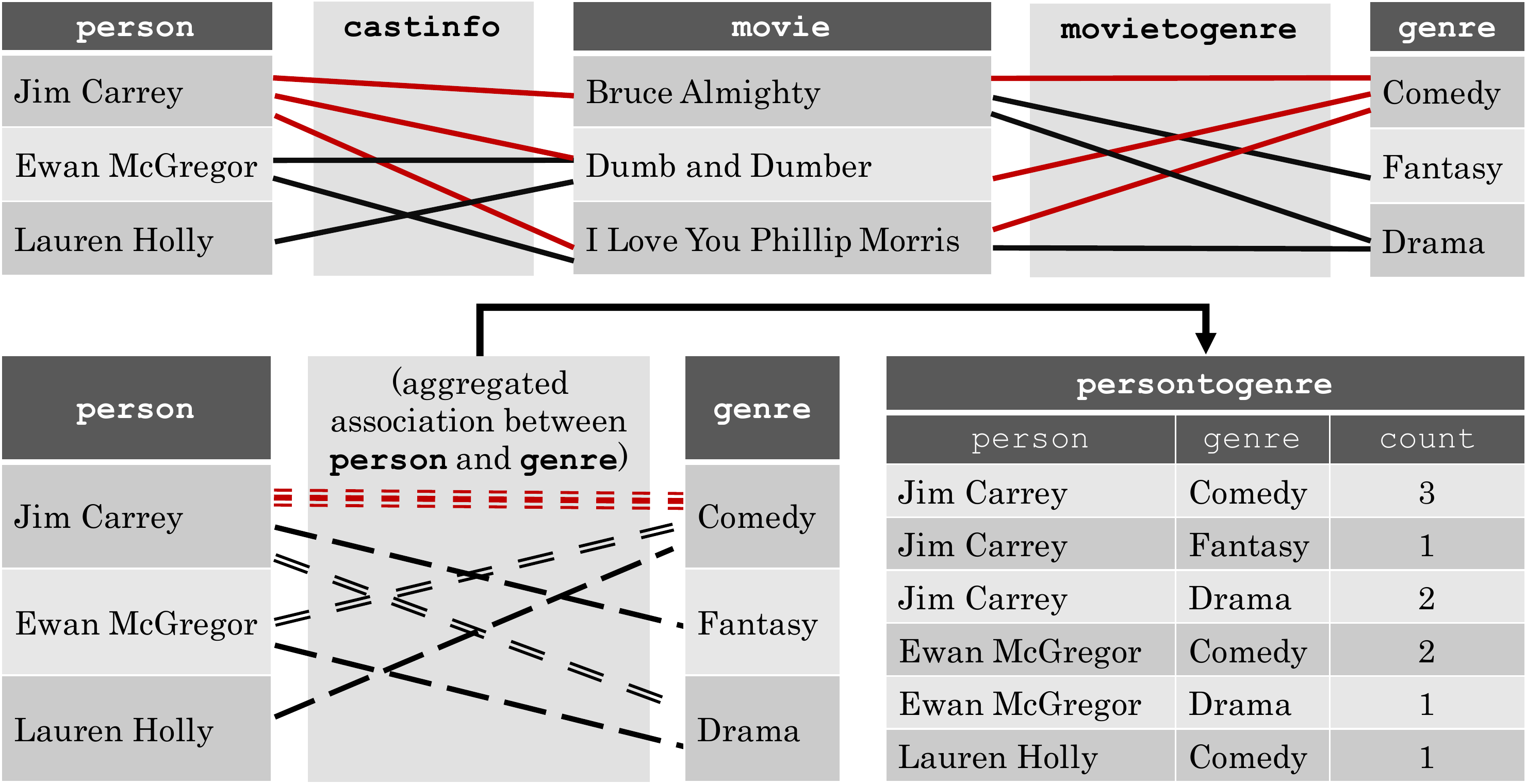}
    \vspace{-5mm}
    \caption{ A \sql{genre} value (e.g., \sql{genre=Comedy}) is a basic
    semantic property of a \sql{movie} (through the
    \sql{movietogenre} relation). A \sql{person} is associated with
    \sql{movie} entities (through the \sql{castinfo} relation);
    aggregates of basic semantic properties of movies are \emph{derived
    semantic properties} of \sql{person}, e.g., the number of comedy movies
    a person appeared in. The \adb stores the derived property in the new
    relation \sql{persontogenre}. (For ease of exposition, we depict
    attributes \sql{genre} and \sql{person} instead of \sql{genre.id}
    and \sql{person.id}.)}
\label{derivedSemanticProperty}
\vspace{-2mm}
\end{figure}

\subsection{Solution Sketch}\label{solSketch}

At the core of \approach is an \emph{abduction-ready database}, \adb
(\fig~\ref{systemArchitecture}). The \adb (1)~increases \approach's efficiency by storing precomputed associations and
statistics, and (2)~simplifies the query model by reducing the extended
family of \queryFamily queries on the original database to equivalent SPJ
queries on the \adb.

\begin{example}\label{precomputationExample} The IMDb database has, among
others, relations \sql{person} and \sql{genre} (\fig~\ref{schema}). \approach's
\adb stores a derived semantic property that associates the two entity types in
a new relation, \sql{persontogenre}\sql{(person.id, genre.id, count)}, which
stores how many movies of each genre each person appeared in. \approach derives
this relation through joins with \sql{castinfo} and \sql{movietogenre}, and
aggregation (\fig~\ref{derivedSemanticProperty}). Then, the \queryFamily query
\sql{Q4} (Example~\ref{funnyActorExample}) is equivalent to the simpler SPJ
query \sql{Q5} on the \adb:\\
\small
\noindent
\smallspace \sql{Q5:} \noindent \sql{SELECT person.name\\
\noindent 
\halftab \smallspace \smallspace FROM person, persontogenre, genre\\ 
\noindent 
\halftab \smallspace\smallspace WHERE person.id = persontogenre.person\_id AND\\
\noindent 
\halftab \halftab persontogenre.genre\_id = genre.id AND\\
\noindent 
\halftab \halftab genre.name = `Comedy' AND persontogenre.count >= 40
} 
\normalsize
\end{example}

\noindent
By incorporating aggregations in precomputed, derived relations, \approach can
reduce \queryFamily queries on the original data to SPJ queries on the \adb.
\approach starts by inferring a PJ query, $Q^*$, on the \adb as a query
template; it then augments $Q^*$ with selection predicates, driven by the
semantic similarity of the examples.
\cref{semanticPropertyFilterSection} formalizes \approach's model of
query intent as a combination of the base query $Q^*$ and a set of semantic
property filters. Then, \cref{model} analyzes the probabilistic
abduction model that \approach uses to solve the query intent discovery problem
(Definition~\ref{def:QuID}).

\looseness-1
After the formal models, we describe the system components of \approach.
\cref{offlinePrecomputation} describes the offline module, which is
responsible for making the database abduction-ready, by precomputing semantic
properties and statistics in derived relations. 
\cref{intentDiscovery} describes the query intent discovery
module, which abduces the most likely intent as an SPJ query on the \adb.

\section{Modeling Query Intent}\label{semanticPropertyFilterSection}

\looseness-1 \approach's core task is to infer the proper SPJ query on the
\adb. We model an SPJ query as a pair of a base query and a set of semantic
property filters: $Q^\FilterSubset {=} (Q^*, \FilterSubset)$. The \emph{base
query} $Q^*$ is a project-join query that captures the structural aspect of the
example tuples. \approach can handle examples with multiple attributes, but,
for ease of exposition, we focus on example tuples that contain a single
attribute of a single entity (\sql{name} of \sql{person}).

In contrast to existing approaches that derive PJ queries from example tuples,
the base query in \approach does not need to be minimal with respect to the
number of joins: While a base query on a single relation with projection on the
appropriate attribute (e.g., \sql{Q1} in Example~\ref{collegeDBExample}) would
capture the structure of the examples, the semantic context may rely on other
relations (e.g., \sql{research}, as in \sql{Q2} of
Example~\ref{collegeDBExample}). Thus, \approach considers any number of joins
among \adb relations for the base query, but limits these to key-foreign key
joins.

\looseness-1
We discuss a simple method for deriving the base query in
Section~\ref{sec:abduction}. \approach's core challenge is
to infer $\FilterSubset$, which denotes a set of \emph{semantic property filters}
that are added as conjunctive selection predicates to $Q^*$.
The base query and semantic property filters
for \sql{Q2} of Example~\ref{collegeDBExample} are:\\
\noindent\smallspace \smallspace $Q^* = $ \small\sql{SELECT name FROM academics, research}\\ 
\noindent\smallspace \smallspace \phantom{$Q^*=$} \sql{WHERE research.aid = academics.id}\normalsize\\
\noindent\smallspace \smallspace $\FilterSubset$ = $\{$ \small\sql{research.interest = `data management'}\normalsize$\}$

\subsection{Semantic Properties and Filters}\label{semanticPropFilterSection}
Semantic properties encode characteristics of an entity. We distinguish
semantic properties into two types. 
(1)~A \emph{basic semantic property} is
affiliated with an entity directly. In the IMDb schema of \fig~\ref{schema},
\sql{gender=Male} is a basic semantic property of a \sql{person}. 
(2)~A \emph{derived
semantic property} of an entity is an aggregate over a basic semantic property
of an associated entity. In Example~\ref{precomputationExample}, the number of
movies of a particular genre that a person appeared in is a derived semantic
property for \sql{person}.
We represent a semantic property $p$ of an entity from a relation $R$ as a triple $p=\prop{A, V, \theta}$.  
In this notation, $V$ denotes a
value\footnote{\approach can support disjunction for categorical attributes
(e.g., \sql{gender=Male or gender=Female}), so $V$ could be a set of values.
However, for ease of exposition we keep our examples limited to properties without
disjunction.} or a value range for attribute $A$ associated with entities in
$R$. The \emph{association strength} parameter $\theta$ quantifies how strongly
an entity is associated with the property. It corresponds to a threshold on
derived semantic properties (e.g., the number of comedies an actor appeared in); it is not defined for basic properties
($\theta=\bot$).

\begin{sloppy}
A \emph{semantic property filter} $\phi_p$ is a structured language
representation of the semantic property $p$.
In the data of \fig~\ref{exampleRunning}, the filters
$\phi_{\prop{{\mathtt{gender}},{\mathtt{Male}},\bot}}$ and $\phi_{\prop{{\mathtt{age}},
[{\mathtt{50, 90}}],\bot}}$ represent two basic semantic properties on
\sql{gender} and \sql{age}, respectively.
Expressed in relational
algebra, filters on basic semantic properties map to standard selection
predicates, e.g.,
$\sigma_{\mathtt{gender} = \mathtt{Male}}$(\sql{person}) and 
$\sigma_{\mathtt{50}\leq\mathtt{age} \leq \mathtt{90}}$(\sql{per\-son}).
For derived properties, filters specify conditions on the association
across different entities. In Example~\ref{precomputationExample}, for
\sql{person} entities, the filter
$\phi_{\prop{{\mathtt{genre}},{\mathtt{Comedy}}, {\mathtt{30}}}}$ denotes the
property of a person being associated with at least 30 movies with the basic
property \sql{genre}=\sql{Comedy}. In relational algebra,
filters on derived properties map to selection predicates over derived
relations in the \adb, e.g., $\sigma_{{\mathtt{genre}} ={\mathtt{Comedy}}\wedge
{\mathtt{count}} \ge {\mathtt{30}}}$(\sql{person\-togenre}).
\end{sloppy}

\subsection{Filters and Example Tuples}\label{sec:filterAndExampleTuples}
To construct $Q^\FilterSubset$, \approach needs to infer the proper set of semantic
property filters given a set of example tuples. Since all example tuples should
be in the result of $Q^\FilterSubset$, $\FilterSubset$ cannot contain filters that the
example tuples do not satisfy. Thus, we only consider \emph{valid} filters that
map to selection predicates that all examples satisfy.

\setlength{\textfloatsep}{0pt} 
\begin{definition}[Filter validity]
    Given a database $\mathcal{D}$, an example tuple set $E$, and a base query
    $Q^*$, a filter $\phi$ is valid if and only if $Q^{\{\phi\}}(\mathcal{D})\supseteq
    E$, where $Q^{\{\phi\}}=(Q^*,\{\phi\})$.
\end{definition}

\fig~\ref{exampleRunning} shows a set of example tuples over the relation
\sql{person}. Given the base query $Q^*{=}$\sql{SELECT name FROM
person}, the filters $\phi_{\prop{{\mathtt{gender}}, {\mathtt{Male}}, \bot}}$ and
$\phi_{\prop{{\mathtt{age}}, {\mathtt{[50,90]}}, \bot}}$ on relation
\sql{person} are valid, because all of the example entities of
\fig~\ref{exampleRunning} are \sql{Male} and fall in the age range [50, 90].

\begin{lemma}\label{compoundFilterLemma} (Validity of conjunctive filters). The
conjunction ($\phi_1\wedge\phi_2\wedge\dots$) of a set of filters
$\varPhi=\{\phi_1,\phi_2,\dots\}$ is valid, i.e.,
$Q^\varPhi(\mathcal{D})\supseteq E$, if and only if $\forall\phi_i\in \varPhi$
$\phi_i$ is valid. \end{lemma}

Relaxing a filter (loosening its conditions) preserves validity. For example,
if $\phi_{\prop{\mathtt{age}, \mathtt{[50,90]}, \bot}}$ is valid, then
$\phi_{\prop{\mathtt{age}, \mathtt{[40,120]}, \bot}}$ is also valid. Out of all
valid filters, \approach focuses on \emph{minimal} valid filters, which have
the tightest bounds.\footnote{Bounds can be derived in different ways,
potentially informed by the result set cardinality. However, we found that the
choice of the tightest bounds works well in practice.}

\begin{definition}[Filter minimality]
    A basic semantic property filter $\phi_{\prop{A,V,\bot}}$ is \emph{minimal}
    if it is valid, and $\forall V^\prime {\subset} V$,
    $\phi_{\prop{A,V^\prime,\bot}}$ is not valid.
    A derived semantic property filter $\phi_{\prop{A,V,\theta}}$ is
    \emph{minimal} if it is valid, and $\forall \epsilon > 0,
    \phi_{\prop{A,V,\theta+\epsilon}}$ is not valid.

\end{definition}

In the example of \fig~\ref{exampleRunning}, $\phi_{\prop{\mathtt{age},
[\mathtt{50,90}],\bot}}$ is a minimal filter and $\phi_{\prop{\mathtt{age},
[\mathtt{40,90}],\bot}}$ is not.

\section{Probabilistic Abduction Model}\label{model}

We now revisit the problem of Query Intent Discovery
(Definition~\ref{def:QuID}), and recast it based on our model of query
intent (\cref{semanticPropertyFilterSection}). Specifically,
Definition~\ref{def:QuID} aims to discover an {\queryFamily} query $Q$; this is
reduced to an equivalent SPJ query $Q^\FilterSubset$ on the \adb (as in
Example~\ref{precomputationExample}). \approach's task is to find the query
$Q^\FilterSubset$ that maximizes the posterior probability
$Pr(Q^\FilterSubset|E)$, for a given set $E$ of example tuples. In this
section, we analyze the probabilistic model to compute this posterior, and
break it down to three components.

\begin{figure}[t]       
    \begin{subtable}{0.28\textwidth}
	\setlength\tabcolsep{1.5pt}
    \centering
    \renewcommand{\arraystretch}{0.95}
    {\small
        \begin{tabular}{|c|l|l|c|}
            \multicolumn{4}{c}{\sql{person}}\\
            \hline
            \rowcolor{vlightgray}
            \multicolumn{1}{|c|}{id} & 
            \multicolumn{1}{c|}{name}&
            \multicolumn{1}{c|}{gender}&
            \multicolumn{1}{c|}{age} \\
            \hline
                1 & Tom Cruise & Male & 50 \\ 
                2 & Clint Eastwood & Male & 90 \\
                3 & Tom Hanks & Male & 60 \\
                4 & Julia Roberts & Female & 50 \\
                5 & Emma Stone & Female & 29 \\
                6 & Julianne Moore & Female & 60 \\
            \hline
        \end{tabular}
	}
    \end{subtable}
    \begin{subtable}{0.15\textwidth}
	\setlength\tabcolsep{1.5pt}
        \centering
    \renewcommand{\arraystretch}{0.95}
    {\small
        \begin{tabular}{|l|}
            \multicolumn{1}{c}{Example tuples}\\
            \hline
            \rowcolor{vlightgray}
            \multicolumn{1}{|c|}{Column 1}\\
            \hline
            % \hline
                Tom Cruise\\ 
                Clint Eastwood\\
            \hline
        \end{tabular}
	}
    \end{subtable}
    \vspace{-2mm}
    \caption{Sample database with example tuples}
    \label{exampleRunning}
    \vspace{2mm}
\end{figure}

\subsection{Notations and Preliminaries}\label{sec:notation}

\noindent\textbf{Semantic context $\bm{\x}$}. Observing a semantic property in
a set of 10 examples is more significant than observing the same property in a
set of 2 examples. We denote this distinction with the \emph{semantic context}
$\x = (p, |E|)$, which encodes the size of the set ($|E|$) where the semantic
property $p$ is observed. We denote with $\mathcal{\X}=\{\x_1, x_2,\dots\}$ the
set of semantic contexts exhibited by the set of example tuples $E$.
\noindent \textbf{Candidate SPJ query $\bm{Q^\FilterSubset}$}. Let
$\allFilterSet = \{\phi_1,\phi_2,\dots\}$ be the set of minimal valid
filters\footnote{We omit $\prop{A,V,\theta}$ in the filter notation when the
context in clear.}, from hereon simply referred to as filters, where $\phi_i$
encodes the semantic context $\x_i$. Our goal is to identify the subset of
filters in $\allFilterSet$ that best captures the query intent. A set of
filters $\FilterSubset\subseteq\allFilterSet$ defines a candidate query
$Q^\FilterSubset=(Q^*, \FilterSubset)$, and
$Q^\FilterSubset(\mathcal{D})\supseteq E$ (from
Lemma~\ref{compoundFilterLemma}).

\noindent \textbf{Filter event $\widetilde{\bm{\phi}}$}. A filter
$\phi\in\allFilterSet$ may or may not appear in a candidate query
$Q^\FilterSubset$. With slight abuse of notation, we denote the filter's
presence ($\phi\in\FilterSubset$) with $\phi$ and its absence
($\phi\not\in\FilterSubset$) with $\bar\phi$. We use $\widetilde{\phi}$ to
represent the occurrence event of $\phi$ in $Q^\FilterSubset$.

Thus:
$\widetilde{\phi} = \left\{\def\arraystretch{1.2}\begin{tabular}{@{}l@{\quad}l@{}}
  $\phi$ & if $\phi\in\FilterSubset$ \\
  $\bar\phi$ & if $\phi \not\in \FilterSubset$
\end{tabular}\right.$

\begin{figure}
	\centering
		{\small
        \begin{tabular}{|p{2.25cm}|l|}
			\hline
			\rowcolor{vlightgray}
            \multicolumn{1}{|c|}{Notation} & \multicolumn{1}{c|}{Description}\\
			\hline
   			$p=\prop{A, V, \theta}$ & Semantic property defined by attribute $A$,\\& value $V$, and 
			association strength $\theta$\\\hline
			$\phi_p$ or $\phi$ & Semantic property filter for $p$\\\hline
			$\allFilterSet  = \{\phi_1, \phi_2, \dots \}$ & Set of minimal valid filters\\\hline						
			$Q^{\FilterSubset} = (Q^*, \FilterSubset)$ & SPJ query with semantic property filters 
														 $\FilterSubset \subseteq \allFilterSet$\\& applied
														 on base query $Q^*$\\\hline
			$x = (p, |E|)$ & Semantic context of $E$ for $p$\\\hline
			$\mathcal{X} = \{x_1, x_2, \dots\}$ & Set of semantic contexts\\
			\hline
        \end{tabular}
		}
		\vspace{-2mm}
	\caption{Summary of notations}
	\label{tab:notations}
	\vspace{2mm}
\end{figure}
\subsection{Modeling Query Posterior} We first analyze the probabilistic model
for a \emph{fixed base query} $Q^*$ and then generalize the model in
Section~\ref{sec:generalize}. We use $Pr_*(a)$ as a shorthand for $Pr(a|Q^*)$. We
model the query posterior $Pr_*(Q^\FilterSubset|E)$, using Bayes' rule:
\begin{align}
    \nonumber 
    Pr_*(Q^\FilterSubset|E) = \frac{Pr_*(E|Q^\FilterSubset)Pr_*(Q^\FilterSubset)}{Pr_*(E)}
\end{align}

By definition, $Pr_*(\mathcal{\X}|E) = 1$; therefore:
\begin{align}
    \nonumber 
    Pr_*(Q^\FilterSubset|E) 
    &= \frac{Pr_*(E, \mathcal{\X}|Q^\FilterSubset)Pr_*(Q^\FilterSubset)}{Pr_*(E)}\\
    \nonumber
    &= \frac{Pr_*(E|\mathcal{\X},Q^\FilterSubset) Pr_*(\mathcal{\X}|Q^\FilterSubset)Pr_*(Q^\FilterSubset)}{Pr_*(E)}
\end{align}

Using the fact that $Pr_*(\mathcal{\X}|E) = 1$ and applying Bayes' rule on the prior $Pr_*(E)$, we get:
\begin{align}
\nonumber Pr_*(Q^\FilterSubset|E) 
=\frac{Pr_*(E|\mathcal{\X},Q^\FilterSubset) Pr_*(\mathcal{\X}|Q^\FilterSubset)Pr_*(Q^\FilterSubset)}{Pr_*(E|\mathcal{\X})Pr_*(\mathcal{\X})}
\end{align}

Finally, $E$ is conditionally independent of $Q^\FilterSubset$ given
the semantic context $\mathcal{\X}$, i.e., $Pr_*(E|\mathcal{\X},
Q^\FilterSubset){=}Pr_*(E|\mathcal{\X})$. Thus:
\begin{align}
	Pr_*(Q^\FilterSubset|E) = \frac{Pr_*(\mathcal{\X}|Q^\FilterSubset)Pr_*(Q^\FilterSubset)}{Pr_*(\mathcal{\X})}\label{qpEqn}
\end{align}

\noindent In \eqn~\ref{qpEqn}, we have modeled the query posterior in terms
of~three components: (1)~the semantic context prior $Pr_*(\mathcal{\X})$,
(2)~the qu\-ery prior $Pr_*(Q^\FilterSubset)$, and (3)~the semantic context
posterior, $Pr_*(\mathcal{\X}|Q^\FilterSubset)$. We proceed to analyze each of
these components.

\subsubsection{Semantic Context Prior}\label{contextPrior}

The semantic context prior $Pr_*(\mathcal{\X})$ denotes the probability
that any set of of example tuples of size $|E|$ exhibits the semantic contexts
$\mathcal{\X}$. This probability is not easy to compute analytically, as it
involves computing a marginal over a potentially infinite set of candidate
queries. In this work, we model the semantic context prior as proportional to
the \emph{selectivity} $\psi(\allFilterSet)$ of
$\allFilterSet=\{\phi_1,\phi_2,\dots\}$, where $\phi_i \in \allFilterSet$ is a
filter that encodes context $\x_i \in \mathcal{\X}$:
\begin{align}
	\label{semanticContextPriorEqn}
	Pr_*(\mathcal{\X}) \propto \psi(\allFilterSet)
\end{align}

\noindent
\textbf{Selectivity $\bm{\psi(\phi)}$}. 
The selectivity of a filter $\phi$ denotes the portion of tuples from the
result of the base query $Q^*$ that satisfy $\phi$:
\vspace{-2mm}
\begin{align*}
    \psi(\phi) = \frac{|Q^{\{\phi\}}(\mathcal{D})|}{|Q^*(\mathcal{D})|}
\end{align*}

Similarly, for a set of filters $\allFilterSet$, $\psi(\allFilterSet) =
\frac{|Q^{\allFilterSet}(\mathcal{D})|}{|Q^*(\mathcal{D})|}$. Intuitively, a
selectivity value close to $1$ means that the filter is not very selective and
most tuples satisfy the filter; selectivity value close to $0$ denotes that
the filter is highly selective and rejects most of the tuples. For example, in
\fig~\ref{exampleRunning}, $\phi_{\prop{\mathtt{gender}, \mathtt{Male},\bot}}$
is more selective than $\phi_{\prop{\mathtt{age},[\mathtt{50,90}],\bot}}$,
with selectivities $\frac{1}{2}$ and $\frac{5}{6}$, respectively.

Selectivity captures the rarity of a semantic context: uncommon contexts are
present in fewer tuples and thus appear in the output of fewer queries.
Intuitively, a rare context has lower prior probability of being observed,
which supports the assumption of \eqn~\ref{semanticContextPriorEqn}.

\subsubsection{Query Prior}\label{queryPriorSection}

The query prior $Pr_*(Q^\FilterSubset)$ denotes the probability that
$Q^\FilterSubset$ is the intended query, prior to observing the example tuples.
We model the query prior as the joint probability of all filter events
$\widetilde{\phi}$, where ${\phi}\in \allFilterSet$. By further assuming filter
independence\footnote{Reasoning about database queries commonly assumes
independence across selection predicates, which filters represent, even though
it may not hold in general.}, we reduce the query prior to a product of
probabilities of filter events:
\begin{align}\label{queryPrior}
	 Pr_*(Q^\FilterSubset) = Pr_*(\textstyle\bigcap_{\phi \in \allFilterSet}\widetilde{\phi}) =  \textstyle\prod_{\phi \in \allFilterSet} Pr_*(\widetilde{\phi})
\end{align}

The \emph{filter event prior} $Pr_*(\widetilde{\phi})$ denotes the prior
probability that filter $\phi$ is included in (if $\widetilde{\phi}=\phi$) or
excluded from (if $\widetilde{\phi}=\bar\phi$) the intended query. We compute
$Pr_*(\widetilde{\phi})$ for each filter as follows:
\begin{align*} 
Pr_*(\phi) = \rho \cdot \delta(\phi) \cdot \alpha(\phi) \cdot \lambda(\phi) 
\;\text{ and }\;
Pr_*(\bar\phi) = 1 - Pr_*(\phi)
\end{align*}
Here, $\rho$ is a base prior parameter, common across all filters, and
represents the default value for the prior. The other factors ($\delta$,
$\alpha$, and $\lambda$) reduce the prior, depending on
characteristics of each filter. We describe these parameters next.

\noindent \looseness-1
\textbf{Domain selectivity impact $\bm{\delta(\phi)}$.} 
Intuitively, a filter that covers a large range of values in an attribute's
domain is unlikely to be part of the intended query. For example, if a user is
interested in actors of a certain age group, that age group is more likely to
be narrow ($\phi_{\prop{\mathtt{age,[41, 45]},\bot}}$) than broad
($\phi_{\prop{\mathtt{age,[41, 90]},\bot}}$). We penalize broad filters with
the parameter $\delta\in(0,1]$; $\delta(\phi)$ is equal to 1 for filters that
do not exceed a predefined ratio in the coverage of their domain, and decreases
for filters that exceed this threshold.\footnote{{Details on the computation of
$\delta(\phi)$ and $\lambda(\phi)$ are in \ifTechRep the Appendix. \else
\theTechRep. \fi}}

\noindent 
\textbf{Association strength impact $\bm{\alpha(\phi)}$.} 
Intuitively, a derived filter with low association strength is unlikely to
appear in the intended query, as the filter denotes a weak association with the
relevant entities. For example,
$\phi_{\prop{\mathtt{genre},\mathtt{Comedy},\mathtt{1}}}$ is less likely than
$\phi_{\prop{\mathtt{genre},\mathtt{Comedy},\mathtt{30}}}$ to represent a query
intent. We label filters with $\theta$ lower than a threshold $\tau_\alpha$ as
insignificant, and set $\alpha(\phi)=0$. All other filters, including basic
filters, have $\alpha(\phi)=1$.

\noindent \looseness-1
\textbf{Outlier impact $\bm{\lambda(\phi)}$.} 
While $\alpha(\phi)$ characterizes the impact of association strength on a
filter individually, $\lambda(\phi)$ characterizes its impact in consideration
with other derived filters over the same attribute. \fig~\ref{skewnessOutlier}
demonstrates two cases of derived filters on the same attribute
(\sql{genre}), corresponding to two different sets of example tuples. In
Case A, $\phi_1$ and $\phi_2$ are more significant than the other filters of
the same family (higher association strength). Intuitively, this corresponds to
the intent to retrieve actors who appeared in mostly \sql{Comedy} and
\sql{SciFi} movies. In contrast, Case B does not have filters that stand
out, as all have similar association strengths: The actors in
this example set are not strongly associated with particular genres, and thus,
intuitively, this family of filters is not relevant to the query intent.

\begin{figure}[t]   
    \begin{subtable}{0.23\textwidth}
    \centering
	\renewcommand{\arraystretch}{0.85}
		{\small
        \begin{tabular}{|r|rll|}
            \hline
			\rowcolor{vlightgray}

            \multicolumn{4}{|c|}{Case A}\\
			
            \hline
            \hline
                $\phi_1\!\!$ & $\!\!\phi_{\langle\mathtt{genre},\!\!\!\!}$ & $\phantom{}_{\mathtt{Comedy}}$,&$_{\mathtt{30 }\rangle}$\\
                $\phi_2\!\!$ & $\!\!\phi_{\langle\mathtt{genre},\!\!\!\!}$ & $\phantom{}_{\mathtt{SciFi}}$,&$_{\mathtt{25 }\rangle}$\\
                $\phi_3\!\!$ & $\!\!\phi_{\langle\mathtt{genre},\!\!\!\!}$ & $\phantom{}_{\mathtt{Drama}}$,&$_{\mathtt{ 3 }\rangle}$\\
                $\phi_4\!\!$ & $\!\!\phi_{\langle\mathtt{genre},\!\!\!\!}$ & $\phantom{}_{\mathtt{Action}}$,&$_{\mathtt{ 2 }\rangle}$\\
                $\phi_5\!\!$ & $\!\!\phi_{\langle\mathtt{genre},\!\!\!\!}$ & $\phantom{}_{\mathtt{Thriller}}$,&$_{\mathtt{ 1 }\rangle}$\\
            \hline
        \end{tabular}
		}
    \end{subtable}
	\hspace{1mm}
    \begin{subtable}{0.23\textwidth}
    \centering
	\renewcommand{\arraystretch}{0.85}
		{\small
        \begin{tabular}{|r|rll|}
            \hline
			\rowcolor{vlightgray}
            \multicolumn{4}{|c|}{Case B}\\
            \hline
            \hline
                $\phi_1\!\!$ & $\!\!\phi_{\langle\mathtt{genre},\!\!\!\!}$ & $\phantom{}_{\mathtt{Comedy}}$,&$_{\mathtt{12 }\rangle}$\\
                $\phi_2\!\!$ & $\!\!\phi_{\langle\mathtt{genre},\!\!\!\!}$ & $\phantom{}_{\mathtt{SciFi}}$,&$_{\mathtt{10 }\rangle}$\\
                $\phi_3\!\!$ & $\!\!\phi_{\langle\mathtt{genre},\!\!\!\!}$ & $\phantom{}_{\mathtt{Drama}}$,&$_{\mathtt{10 }\rangle}$\\
                $\phi_4\!\!$ & $\!\!\phi_{\langle\mathtt{genre},\!\!\!\!}$ & $\phantom{}_{\mathtt{Action}}$,&$_{\mathtt{ 9 }\rangle}$\\
                $\phi_5\!\!$ & $\!\!\phi_{\langle\mathtt{genre},\!\!\!\!}$ & $\phantom{}_{\mathtt{Thriller}}$,&$_{\mathtt{ 9 }\rangle}$\\
            \hline
        \end{tabular}
		}
    \end{subtable}
	
	\vspace{-1mm}
    \caption{Two cases for derived semantic property filters. Top two filters
    of Case A are interesting, whereas no filter is interesting in Case B.}
\label{skewnessOutlier}
\vspace{2mm}
\end{figure}

\setcounter{footnote}{\value{footnote}-1}

We model the outlier impact $\lambda(\phi)$ of a filter using the
\emph{skewness} of the association strength distribution within the family of
derived filters sharing the same attribute. Our assumption is that
highly-skewed, heavy-tailed distributions (Case A) are likely to contain the
significant (intended) filters as \emph{outliers}. We set $\lambda(\phi)=1$ for
a derived filter whose association strength is an outlier in the association
strength distribution of filters of the same family. We also set
$\lambda(\phi)=1$ for basic filters. All other filters get
$\lambda(\phi)=0$.\footnotemark{}

\subsubsection{Semantic Context Posterior} 

The semantic context posterior $Pr_*(\mathcal{\X}|Q^\FilterSubset)$ is
the probability that a set of example tuples of size $|E|$, sampled from the
output of a particular query $Q^\FilterSubset$, exhibits the set of semantic
contexts $\mathcal{\X}$:
\begin{align*}
Pr_*(\mathcal{\X}|Q^\FilterSubset)=Pr_*(\x_1,\x_2,...,\x_n|Q^\FilterSubset)
\end{align*}
Two semantic contexts $\x_i, \x_j \in \mathcal{\X}$ are conditionally
independent given $Q^\FilterSubset$. Therefore:
\begin{align*}
Pr_*(\mathcal{\X}|Q^\FilterSubset)=\textstyle\prod_{i=1}^nPr_*(\x_i|Q^\FilterSubset)=\prod_{i=1}^nPr_*(\x_i|\widetilde{\phi}_1, \widetilde{\phi}_2,\dots)
\end{align*}
Recall that $\phi_i$ encodes the semantic context $x_i$ (\cref{sec:notation}).
We assume that ${\x_{i}}$ is conditionally independent of any $\widetilde{\phi}_j, i
\neq j$, given $\widetilde{\phi}_i$ (this always holds for $\widetilde{\phi}_i=\phi_i$):
\begin{align}
Pr_*(\mathcal{\X}|Q^\FilterSubset)=\textstyle\prod_{i=1}^nPr_*(\x_i|\widetilde{\phi}_i)
\label{semanticContextPosterior}
\end{align}

For each $x_i$, we compute $Pr_*({\x_i}|\widetilde{\phi}_i)$ based on the
state of the filter event ($\widetilde{\phi}_i=\phi_i$ or
$\widetilde{\phi}_i=\bar\phi_i$):

\noindent $\bm{Pr_*(\x_i|\phi_i)}$: By definition, all tuples in
$Q^{\{\phi_i\}}(\mathcal{D})$ exhibit the property of $\x_i$. Hence,
$Pr_*(\x_i|\phi_i) = 1$.

\noindent $\bm{Pr_*(\x_i|\bar{\phi}_i)}$: This is the probability that a set of
$|E|$ tuples drawn uniformly at random from $Q^*(\mathcal{D})$ ($\phi_i$ is not
applied to the base query) exhibits the context $\x_i$. The portion of tuples
in $Q^*(\mathcal{D})$ that exhibit the property of $\x_i$ is the selectivity
$\psi(\phi_i)$. Therefore, $Pr_*(\x_i|\bar\phi_i) \approx \psi(\phi_i)^{|E|}$.

\smallskip

Using \eqns~\eqref{qpEqn}--\eqref{semanticContextPosterior}, we derive
the final form of the query posterior (where $K$ is a normalization constant):
\begin{align}
\nonumber
&\resizebox{0.77\hsize}{!}{
$\displaystyle Pr_*(Q^\FilterSubset|E) =
\frac{K}{\psi(\allFilterSet)} \prod_{\phi_i\in\allFilterSet}\left(Pr_*(\widetilde{\phi}_i)Pr_*({\x_i}|\widetilde{\phi}_i)\right)$
}\\
&\resizebox{0.91\hsize}{!}{
$\displaystyle =
\frac{K}{\psi(\allFilterSet)}\prod_{\phi_i\in\FilterSubset}
Pr_*({\phi}_i)Pr_*({\x_i}|{\phi}_i)
\prod_{\phi_i\not\in\FilterSubset}    Pr_*(\bar{\phi}_i)Pr_*({\x_i}|\bar{\phi}_i)$
}
\label{mainDerivation}
\end{align}

\subsection{Generalization}\label{sec:generalize} 

So far, our analysis focused on
a fixed base query. Given an SPJ query $Q^\FilterSubset$, the underlying base
query $Q^*$ is deterministic, i.e., $Pr(Q^*|Q^\FilterSubset) = 1$. Hence:
\begin{align*}
Pr(Q^\FilterSubset|E) &= Pr(Q^\FilterSubset, Q^* |E)
                = Pr(Q^\FilterSubset|Q^*, E)Pr(Q^*|E)\\
                &= Pr_*(Q^\FilterSubset|E)Pr(Q^*|E)
\end{align*}
We assume $Pr(Q^*|E)$ to be equal for all valid base queries, where
$Q^*(\mathcal{D}) \supseteq E$. Then we use $Pr_*(Q^\FilterSubset|E)$ to find
the query $Q$ that maximizes the query posterior $Pr(Q|E)$.

\section{Offline Abduction Preparation}\label{offlinePrecomputation}

\looseness-1 In this section, we discuss system considerations to perform query
intent discovery \emph{efficiently}. \approach employs an offline module that
performs several pre-computation steps to make the database
\emph{abduction-ready}. The abduction-ready database (\adb) augments the
original database with derived relations that store associations across
entities and precomputes semantic property statistics. Deriving this
information is relatively straightforward; the contributions of this section
lie in the design of the \adb, the information it maintains, and its role in
supporting efficient query intent discovery. We describe the three major
functions of the \adb.

\smallskip

\noindent \textbf{Entity lookup.} \approach's goal is to discover the most
likely query, based on the user-provided examples. To do that, it first needs
to determine which entities in the database correspond to the examples.
\approach uses a \emph{global inverted column index}~\cite{ShenSIGMOD2014},
built over all text attributes and stored in the \adb, to perform fast lookups,
matching the provided example data to entities in the database.

\smallskip

\noindent \textbf{Semantic property discovery.} To reason about intent,
\approach first needs to determine what makes the examples similar. \approach
looks for semantic properties within entity relations (e.g., \sql{gender}
appears in table \sql{person}), other relations (e.g., \sql{genre} appears in a
separate table joining with \sql{movie} through a key-foreign-key constraint),
and other entities, (e.g., the number of movies of a particular \sql{genre}
that a \sql{person} has appeared in). The \adb precomputes and stores such
\emph{derived relations} (e.g., \sql{persontogenre}), as these frequently
involve several joins and aggregations and performing them at runtime would be
prohibitive.\footnote{The data cube~\cite{dataCube} can serve as an alternative
mechanism to model the \adb data, but is much less efficient compared to the
\adb (details are in \ifTechRep Appendix~\ref{dataCube}\else\citeTechRep\fi).}
For example, \approach computes the \sql{persontogenre} relation
(\fig~\ref{derivedSemanticProperty}) and stores it in the \adb with the SQL
query below:

\small
\noindent \smallspace \smallspace  \sql{Q6:}  \sql{CREATE TABLE persontogenre as}\\
\noindent \smallspace \smallspace  \phantom{\sql{Q6:}}  \sql{(SELECT person\_id, genre\_id, count(*) AS count} \\
\noindent \smallspace \smallspace  \phantom{\sql{Q6:}}  \sql{ FROM castinfo, movietogenre}\\
\noindent \smallspace \smallspace  \phantom{\sql{Q6:}}  \sql{ WHERE castinfo.movie\_id = movietogenre.movie\_id}\\
\noindent \smallspace \smallspace  \phantom{\sql{Q6:}}  \sql{ GROUP BY person\_id, genre\_id)}
\normalsize

For the \adb construction, \approach only relies on very basic information to
understand the data organization. It uses (1)~the database schema, including
the specification of primary and foreign key constraints, and (2)~additional
meta-data, which can be provided once by a database administrator, that
specify which tables describe entities (e.g., \sql{person}, \sql{movie}), and
which tables and attributes describe direct properties of entities (e.g.,
\sql{genre}, \sql{age}). \approach then automatically discovers fact tables,
which associate entities and properties, by exploiting the key-foreign key
relationships. \approach also automatically discovers derived properties up to
a certain pre-defined depth, using paths in the schema graph, that connect
entities to properties. Since the number of possible values for semantic
properties is typically very small and remains constant as entities grow, the
\adb grows linearly with the data size. In our implementation, we restrict the
derived property discovery to the depth of two fact-tables (e.g., \approach
derives \sql{persontogenre} through \sql{castinfo} and \sql{movietogenre}).
\approach can support deeper associations, but we found these are not common
in practice. \approach generally assumes that different entity types appear in
different relations, which is the case in many commonly-used schema types,
such as star, galaxy, and fact-constellation schemas. \approach can perform
inference in a denormalized setting, but would not be able to produce and
reason about derived properties in those cases.

\smallskip

\noindent
\textbf{Smart selectivity computation.}
For basic filters involving categorical values, \approach stores the
selectivity for each value. However, for numeric ranges, the number of possible
filters can grow quadratically with the number of possible values. \approach
avoids wasted computation and space by only precomputing selectivities
$\psi(\phi_{\prop{A,[min_{V_A}, v],\bot}})$ for all $v \in V_A$, where $V_A$ is
the set of values of attribute $A$ in the corresponding relation, and $min_{V_A}$ is the minimum value in
$V_A$. The \adb can derive the selectivity of a filter with any value range as:
\begin{align*} 
	\psi(\phi_{\prop{A,(l, h],\bot}}) = \psi(\phi_{\prop{A,[min_{V_A}, h],\bot}}) - \psi(\phi_{\prop{A,[min_{V_A}, l],\bot}}) 
\end{align*}
In case of derived semantic properties, \approach precomputes
selectivities $\psi(\phi_{\prop{A, v, \theta}})$ for all $v \in V_A,
\theta \in \Theta_{A, v}$, where $\Theta_{A, v}$ is the set of values of
association strength for the property ``$A=v$''.

\section{Query Intent Discovery}\label{intentDiscovery}

During normal operation, \approach receives example tuples from a user,
consults the \adb, and infers the most likely query intent
(Definition~\ref{def:QuID}). In this section, we describe how \approach
resolves ambiguity in the provided examples, how it derives their semantic
context, and how it finally abduces the intended query.

\subsection{Entity and Context Discovery}\label{semanticContextDiscovery} 

\approach's probabilistic abduction model (\cref{model}) relies on the set of
semantic contexts $\mathcal{\X}$ and determines which of these
contexts are intended vs coincidental, by the inclusion or exclusion of the
corresponding filters in the inferred query. To derive the set of semantic
contexts from the examples, \approach first needs to
identify the entities in the \adb that correspond to the provided examples.

\subsubsection{Entity disambiguation}\label{sec:entityDisambiguation}

User-provided examples are not complete tuples, but often single-column values
that correspond to an entity. As a result, there may be ambiguity that
\approach needs to resolve. For example, suppose the user provides the
examples: \sql{\{Titanic, Pulp Fiction, The Matrix\}}. \approach consults the
precomputed inverted column index to identify the attributes
(\sql{movie.title}) that contain all the example values, and classifies the
corresponding entity (\sql{movie}) as a potential match. However, while the
dataset contains unique entries for \sql{Pulp Fiction (1994)} and \sql{The
Matrix (1999)}, there are 4 possible mappings for \sql{Titanic}: (1)~a 1915
Italian film, (2)~a 1943 German film, (3)~a 1953 film by Jean Negulesco, and
(4)~the 1997 blockbuster film by James Cameron.

\looseness-1
The key insight for resolving such ambiguities is that the provided examples
are more likely to be alike. \approach selects the entity mappings that
maximize the semantic similarities across the examples. Therefore, based on
the year and country information, it determines that \sql{Titanic} corresponds
to the 1997 film, as it is most similar to the other two (unambiguous)
entities. In case of derived properties, e.g., nationality of actors appearing
in a film, \approach aims to increase the association strength (e.g., the
number of such actors). Since the examples are typically few, \approach can
determine the right mappings by considering all combinations.

\subsubsection{Semantic context discovery}

Once \approach identifies the right entities, it then explores all the semantic
properties stored in the \adb that match these entities (e.g., \sql{year},
\sql{genre}, etc.). Since the \adb precomputes and stores the derived
properties, \approach can produce all the relevant properties using queries
with at most one join. For each property, \approach produces semantic contexts
as follows:

{
\setlength\emergencystretch{.03\textwidth}
\noindent
\textbf{Basic property on categorical attribute.}  
If all examples in $E$ contain value $v$ for the property of attribute $A$,
\approach produces the semantic context $(\prop{A, v, \bot}, |E|)$. For
example, a user provides three movies: \sql{Dunkirk}, \sql{Logan}, and
\sql{Taken}. The attribute \sql{genre} corresponds to a basic property for
movies, and all these movies share the values, \sql{Action} and~\sql{Thriller},
for this property. \approach generates two semantic contexts:
$(\prop{\mathtt{genre},\mathtt{Action}, \bot}, 3)$ and
$(\prop{\mathtt{genre},\mathtt{Thriller}, \bot}, 3)$.

\noindent
\textbf{Basic property on numerical attribute.}
If $v_{min}$ and $v_{max}$ are the minimum and maximum values, respectively,
that the examples in $E$ demonstrate for the property of attribute $A$,
\approach creates a semantic context on the range $[v_{min}, v_{max}]$:
$(\prop{A, [v_{min}, v_{max}], \bot}, |E|)$. For example, if $E$ contains
three persons with ages 45, 50, and 52, \approach will produce the context
$(\prop{\mathtt{age}, [45,52], \bot}, 3)$.

\noindent
\textbf{Derived property.}
If all examples in $E$ contain value $v$ for the derived property of attribute
$A$, \approach produces the semantic context $(\prop{A, v, \theta_{min}},
|E|)$, where $\theta_{min}$ is the minimum association strength for the value
$v$ among all examples. For example, if $E$ contains two persons who have
appeared in 3 and 5 \sql{Comedy} movies, \approach will produce the context
$(\prop{\mathtt{genre}, \mathtt{Comedy}, 3}, 2)$.
}

\subsection{Query Abduction} \label{sec:abduction}

\approach starts abduction by constructing a base query that captures the
structure of the example tuples. Once it identifies the entity and attribute
that matches the examples (e.g., \sql{person.name}), it forms the minimal PJ
query (e.g., \sql{SELECT name FROM person}). It then iterates through the
discovered semantic contexts and appends the corresponding relations to the
\sql{FROM} clause and the appropriate key-foreign key join conditions in the
\sql{WHERE} clause. Since the \adb precomputes and stores the derived
relations, each semantic context will add at most one relation to the query.

The number of candidate base queries is typically very small. For each base
query $Q^*$, \approach abduces the best set of filters
$\FilterSubset\subseteq\allFilterSet$ to construct SPJ query
$Q^\FilterSubset$, by augmenting the \sql{WHERE} clause of $Q^*$ with the
corresponding selection predicates. (\approach also removes from
$Q^\FilterSubset$ any joins that are not relevant to the selected
filters $\FilterSubset$).

While the number of candidate SPJ queries grows exponentially in the number of
minimum valid filters ($2^{|\allFilterSet|}$), we prove that we can make
decisions on including or excluding each filter independently.
Algorithm~\ref{squidAlgo} iterates over the set of minimal valid filters
$\allFilterSet$ and decides to include a filter only if its addition to the
query increases the query posterior probability
(lines~\ref{ln:comp}-\ref{ln:add}). Our query abduction algorithm has
$O(|\allFilterSet|)$ time complexity and is guaranteed to produce the query
$Q^\FilterSubset$ that maximizes the query posterior.

\begin{algorithm}[t]
  \LinesNumbered
  {\small
 \KwIn{set of entities $E$, base query $Q^*$, set of minimal valid filters $\allFilterSet\!\!\!$}
  \KwOut{$Q^\FilterSubset$ such that $Pr_*(Q^\FilterSubset|E)$ is maximized}
  $\mathcal{\X} = \{\x_1, \x_2, ...\}$ \tcp*[f]{semantic contexts in $E$}\\
  $\FilterSubset = \emptyset$\\
  \ForEach{$\phi_i \in \allFilterSet$}{
        $include_{\phi_i} = Pr_*(\phi_i)Pr_*(\x_i|\phi_i)$
        \tcp*[f]{from \eqn~\eqref{mainDerivation}}
        \label{ln:include}\\
        $exclude_{\phi_i} = Pr_*(\bar\phi_i)Pr_*(\x_i|\bar{\phi_i})$
        \tcp*[f]{from \eqn~\eqref{mainDerivation}}
        \label{ln:exclude}\\
        \If{ $include_{\phi_i} > exclude_{\phi_i}$\label{ln:comp}}{
            $\FilterSubset = \FilterSubset \cup \{{\phi_i}\}$ \label{ln:add}
        }       
  }  
  \Return $Q^\FilterSubset$\\
  }
\caption{QueryAbduction ($E, Q^*, \allFilterSet$)}
\label{squidAlgo}
\end{algorithm}

\begin{theorem}\label{theTheorem} 
Given a base query $Q^*$, a set of examples $E$, and a set of minimal valid
filters $\allFilterSet$, Algorithm~\ref{squidAlgo} returns the query
$Q^\FilterSubset$, where $\FilterSubset \subseteq \allFilterSet$, such that
$Pr_*(Q^\FilterSubset|E)$ is maximized. 
\end{theorem}

\section{Experiments}\label{experiment}

In this section, we present an extensive experimental evaluation of \approach
over three real-world datasets, with a total of 41 benchmark queries of varying
complexities. Our evaluation shows that \approach is scalable and effective,
even with a small number of example tuples. Our evaluation extends to
qualitative case studies over real-world user-generated examples, which
demonstrate that \approach succeeds in inferring the query intent of real-world
users. We further demonstrate that when used as a query-reverse-engineering
system in a closed-world setting \approach outperforms the state-of-the-art.
Finally, we show that \approach is superior to semi-supervised PU-learning in
terms of both efficiency and effectiveness.

\subsection{Experimental Setup}\label{expSetup} 
We implemented \approach in Java and all experiments were run on a 12x2.66
GHz machine with 16GB RAM running CentOS 6.9 with PostgreSQL 9.6.6.

\vspace{-2mm}
\paragraph*{Datasets and benchmark queries} 

\looseness-1 Our evaluation includes three real-world datasets and a total of
41 benchmark queries, designed to cover a broad range of intents and query
structures. We summarize the datasets and queries below and provide detailed
description in \ifTechRep Appendix~\ref{datasetAndBenchmark}. \else
\theTechRep. \fi

\noindent 
\textbf{IMDb (633 MB):} The dataset contains 15 relations with information on
movies, cast members, film studios, etc. We designed a set of 16
benchmark queries ranging the number of joins (1 to 8 relations), the number of
selection predicates (0 to 7), and the result cardinality (12 to 2512 tuples).

\noindent
\textbf{DBLP (22 MB):} We used a subset of the DBLP data~\cite{mendeleyDataset}, 
with 14 relations, and 16 years (2000--2015) of top 81 conference publications.
We designed 5 queries ranging the number of joins (3 to 8 relations), the
number of selection predicates (2 to 4), and the result cardinality (15 to 468
tuples).

\noindent
\textbf{Adult (4 MB):} This is a single relation dataset containing census data
of people and their income brackets. We generated 20 queries, randomizing the
attributes and predicate values, ranging the number of selection predicates (2
to 7) and the result cardinality (8 to 1404 tuples).

\vspace{-2mm}
\paragraph*{Case study data}
We retrieved several public lists (sources listed in \ifTechRep
Appendix~\ref{datasetAndBenchmark}\else \theTechRep \fi) with human-generated
examples, and identified the corresponding intent. For example, a user-created
list of ``115 funniest actors'' reveals a query intent (funny actors), and
provides us with real user examples (the names in the list). We used this
method to design 3 case studies: funny actors (IMDb), 2000s Sci-Fi movies
(IMDb), and prolific database researchers (DBLP).

\vspace{-2mm}
\paragraph*{Metrics} 

We report query discovery time as a metric of efficiency. We measure
effectiveness using precision, recall, and f-score. If $Q$ is the intended
query, and $Q^\prime$ is the query inferred by \approach, precision is computed
as $\frac{Q^\prime(\mathcal{D})\cap Q(\mathcal{D})}{Q^\prime(\mathcal{D})}$ and
recall as $\frac{Q^\prime(\mathcal{D})\cap Q(\mathcal{D})}{Q(\mathcal{D})}$;
f-score is their harmonic mean. We also report the total number of
predicates in the produced queries and compare them with the actual intended
queries. 

\begin{figure}[t]
\centering
\begin{minipage}{.13\textwidth}
	\hspace{100mm}
\begin{subfigure}{1\textwidth}
  \centering
  \includegraphics[trim={6mm 2mm 1mm 0mm}, width=0.98\textwidth]{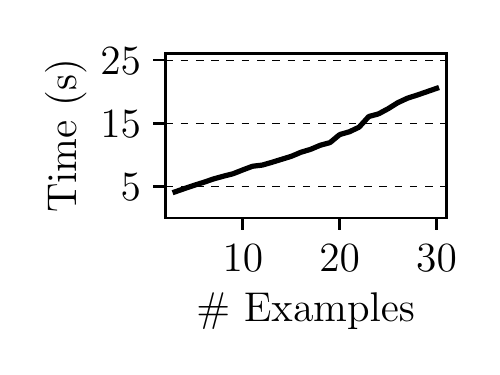}
  \vspace{-7mm}
\end{subfigure}%
\\[1pt]
\begin{subfigure}{1\textwidth}
  \centering
    \includegraphics[trim={4mm 0mm 1mm 3mm}, width=0.98\textwidth]{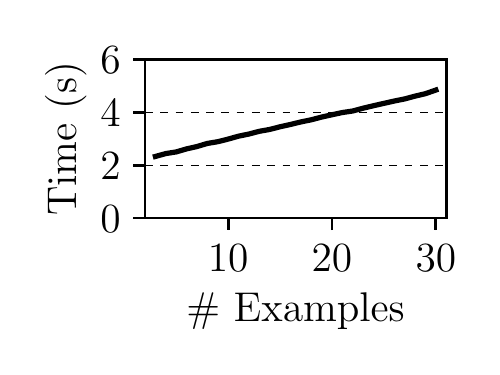}
    \vspace{-7mm}
	\caption{}
    \label{predTime}
	\end{subfigure}
	\vspace{-0mm}
\end{minipage}
\begin{minipage}{.32\textwidth}
\begin{subfigure}{1\textwidth}
    \includegraphics[trim={-10mm 0mm 1mm -5mm}, width=1\textwidth]{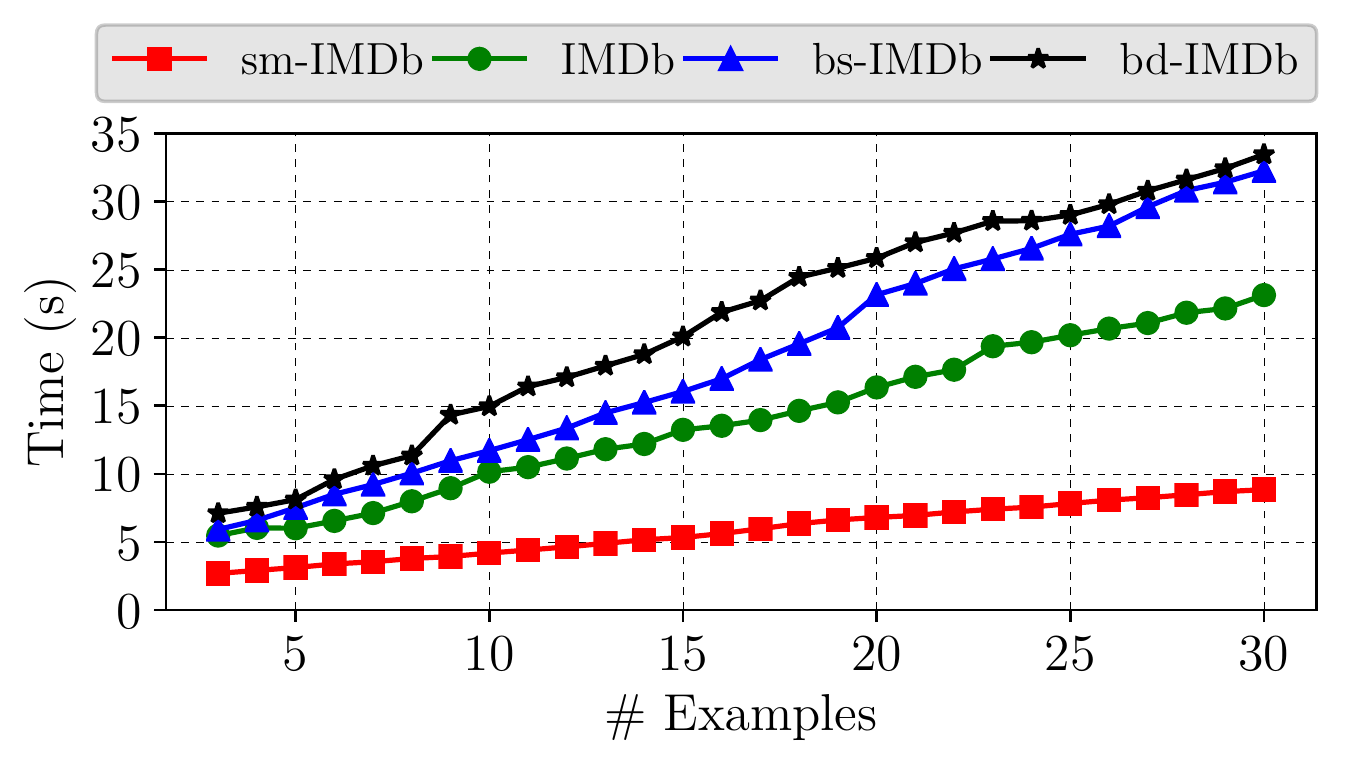}
	\vspace{-6mm}
	\caption{}
    \label{scalabilityFig}
\end{subfigure}
\end{minipage}
    \vspace{-3mm}
    \caption{Average abduction time over the benchmark queries
	 in (a) IMDb (top), DBLP (bottom), and (b) 4 versions of the IMDb 
	 dataset in different sizes.} 
     \vspace{1mm}
\end{figure}

\vspace{-2mm}
\paragraph*{Comparisons} 
To the best of our knowledge, existing QBE techniques do not produce SPJ
queries without (1)~a large number of examples, or (2)~additional information,
such as provenance. For this reason, we can't meaningfully compare \approach
with those approaches. Removing the open-world requirement, \approach is most
similar to the QRE system TALOS~\cite{TranVLDB2014} with respect to
expressiveness and capabilities (Figure~\ref{relatedWorkMatrix}). We compare
the two systems for query reverse engineering tasks in
Section~\ref{qreCompareSection}. We also compare \approach against PU-learning
methods~\cite{elkan2008learning} in Section~\ref{comparePULearning}.

\begin{figure*}[t]
    \centering
	\begin{subfigure}{.71\textwidth}
    \includegraphics[trim={0 4mm 0 0}, width=1\textwidth]{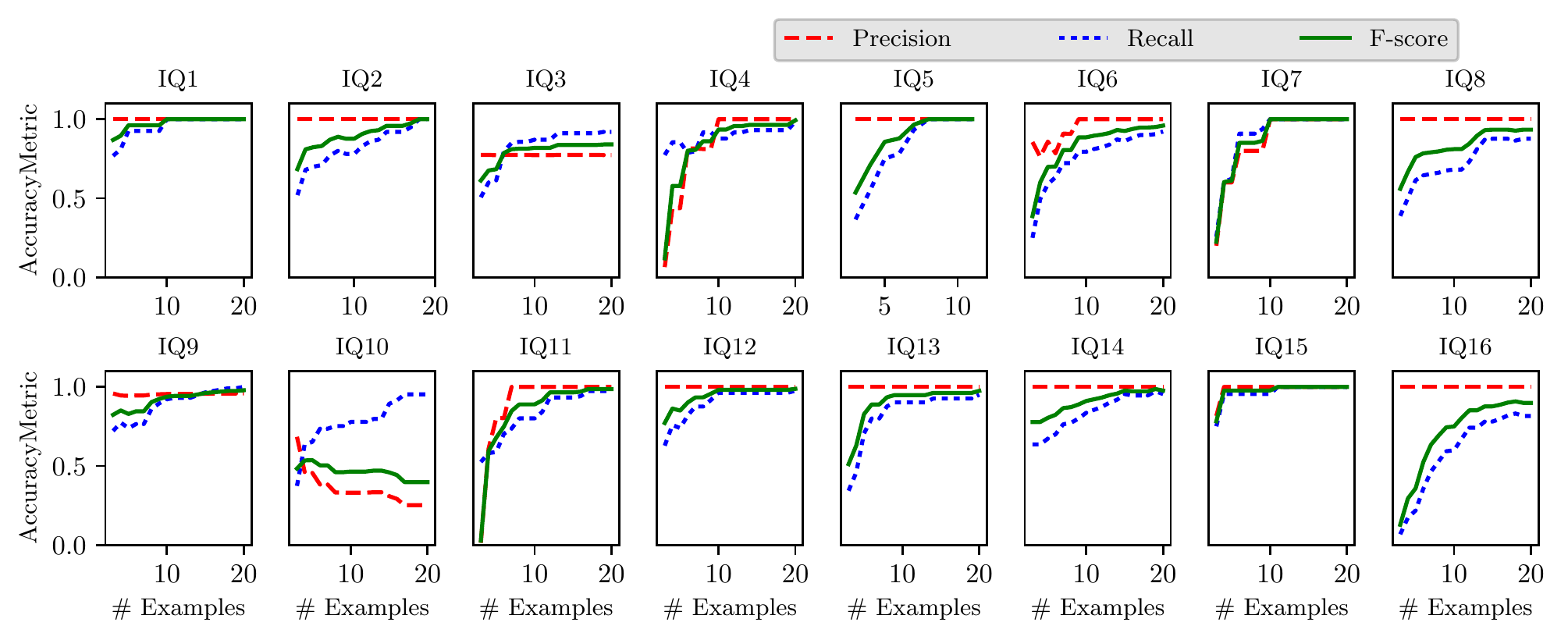}
    \caption{IMDb}
	\vspace{-2mm}
    \label{precisionRecallFscoreIMDb}
	\vspace{-4mm}
	\end{subfigure}
	\rule[-19.5mm]{.3mm}{37mm}
	\begin{subfigure}{.26	\textwidth}
    \includegraphics[trim={2mm 4mm 0 -12mm}, width=1\textwidth]{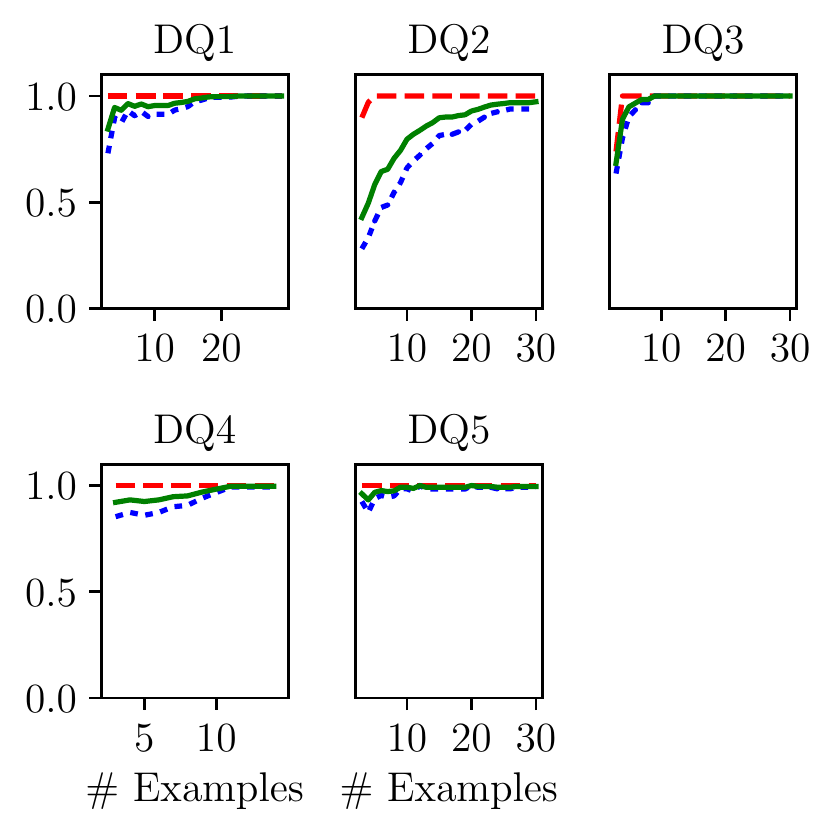}
	\caption{DBLP}
    \label{precisionRecallDblp}
	\vspace{-4mm}
	\end{subfigure}
	\caption{\approach achieves high accuracy with few examples (typically $\sim 5$) in most benchmark queries.}
	\label{precisionRecallFscore}
	\vspace{-2mm}
\end{figure*}

\begin{figure*}[t]
\centering
\begin{minipage}{1\textwidth}
  \begin{subfigure}{.64\textwidth}
  \centering
  \includegraphics[trim={5mm 0mm 0 0}, width=1\textwidth]{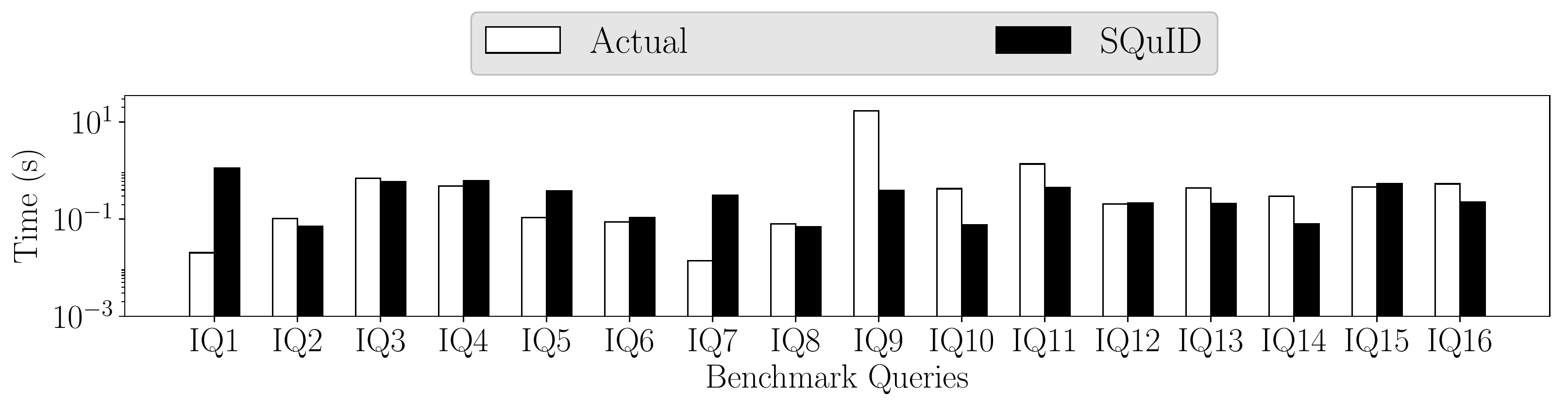}
  \vspace{-6mm}
   \caption{IMDb}
  	\label{execTime}
  \end{subfigure}
  \begin{subfigure}{.35\textwidth}
  \centering
    \includegraphics[trim={0mm 0 0 0}, width=1\textwidth]{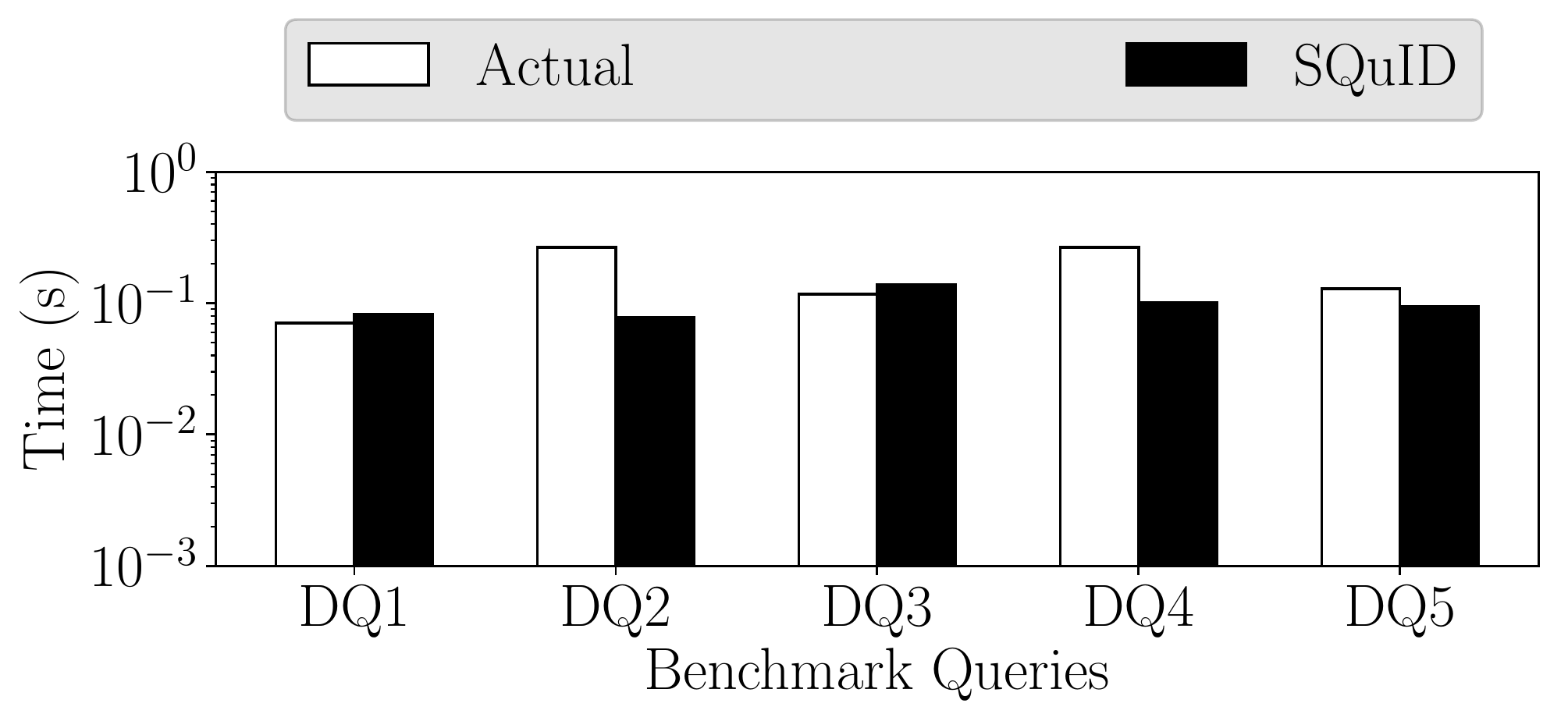}
    \vspace{-6mm}
    \caption{DBLP}
    \label{execTimedblp}
	  \end{subfigure}
\end{minipage}
\vspace{-4mm}
\caption{\approach rarely produces queries that are slower than the original with respect to query runtime.}
\label{comparisonWithOriginalQueries}
\vspace{-4mm}
\end{figure*}

\subsection{Scalability}\label{scalability} 

In our first set of experiments, we examine the scalability of \approach
against increasing number of examples and varied dataset sizes.
\fig~\ref{predTime} displays the abduction time for the IMDb and DBLP datasets
as the number of provided examples increases, averaged over all benchmark
queries in each dataset. Since \approach retrieves semantic properties and
computes context for each example, the runtime increases linearly with the
number of examples, which is what we observe in practice. 

\looseness-1
\fig~\ref{scalabilityFig} extends this experiment to datasets of varied sizes.
We generate three alternative versions of the IMDb dataset: (1)~sm-IMDb {(75
MB)}, a downsized version that keeps 10\% of the original data; (2)~bs-IMDb
{(1330 MB)}, doubles the entities of the original dataset and creates
associations among the duplicate entities (\sql{person} and \sql{movie}) by
replicating their original associations; (3)~bd-IMDb {(1926 MB)}, is the same
as bs-IMDb but also introduces associations between the original entities and
the duplicates, creating denser connections.\footnote{Details of the data
generation process are in \ifTechRep Appendix~\ref{datasetAndBenchmark}\else
\theTechRep\fi.} \approach's runtime increases for all datasets with the number
of examples, and, predictably, larger datasets face longer abduction times.
Query abduction involves point queries to retrieve semantic properties of the
entities, using B-tree indexes. As the data size increases, the runtime of
these queries grows logarithmically. \approach is slower on bd-IMDb than on
bs-IMDb: both datasets include the same entities, but bd-IMDb has denser
associations, which results in additional derived semantic properties.

\subsection{Abduction Accuracy} \label{quality} 

\looseness-1
Intuitively, with a larger number of examples, abduction accuracy should
increase: \approach has access to more samples of the query output, and can
more easily distinguish coincidental from intended similarities.
\fig~\ref{precisionRecallFscore} confirms this intuition, and precision,
recall, and f-score increase, often very quickly, with the number of examples
for most of our benchmark queries.  We discuss here a few particular queries.

\noindent
\sql{IQ4} \& \sql{IQ11}:
These queries include a statistically common property (USA movies), and
\approach needs more examples to confirm that the property is indeed intended,
not coincidental; hence, the precision converges more slowly.

\noindent
\textbf{\sql{IQ6}:}  
In many movies where Clint Eastwood was a director, he was also an actor.
\approach needs to observe sufficient examples to discover that the property
\sql{role:Actor} is not intended, so recall converges more slowly.

\noindent
\textbf{\sql{IQ10}:}
\approach performs poorly for this query. The query looks for actors appearing
in more than 10 \emph{Russian} movies that were released \emph{after 2010}.
While \approach discovers the derived properties ``more than 10 Russian
movies'' and ``more than 10 movies released after 2010'', it cannot compound
the two into ``more than 10 Russian movies released after 2010''. This query
is simply outside of \approach's search space, and \approach produces a query
with more general predicates than was intended, which is why precision drops.

\noindent
\textbf{\sql{IQ3}:}
The query is looking for actresses who are Canadian and were born after 1970.
\approach successfully discovers the properties \sql{gender:Female},
\sql{country:Canada}, and \sql{birth year} $\ge$ \sql{1970}; however, it fails to
capture the property of ``being an actress'', corresponding to having appeared
in at least 1 film. The reason is that \approach is programmed to ignore weak
associations (a person associated with only 1 movie). This behavior can be
fixed by adjusting the association strength parameter to allow for weaker
associations.

\begin{figure}[t]
    \begin{center}
    \includegraphics[trim={4mm 0mm 0 0},width=.48\textwidth]{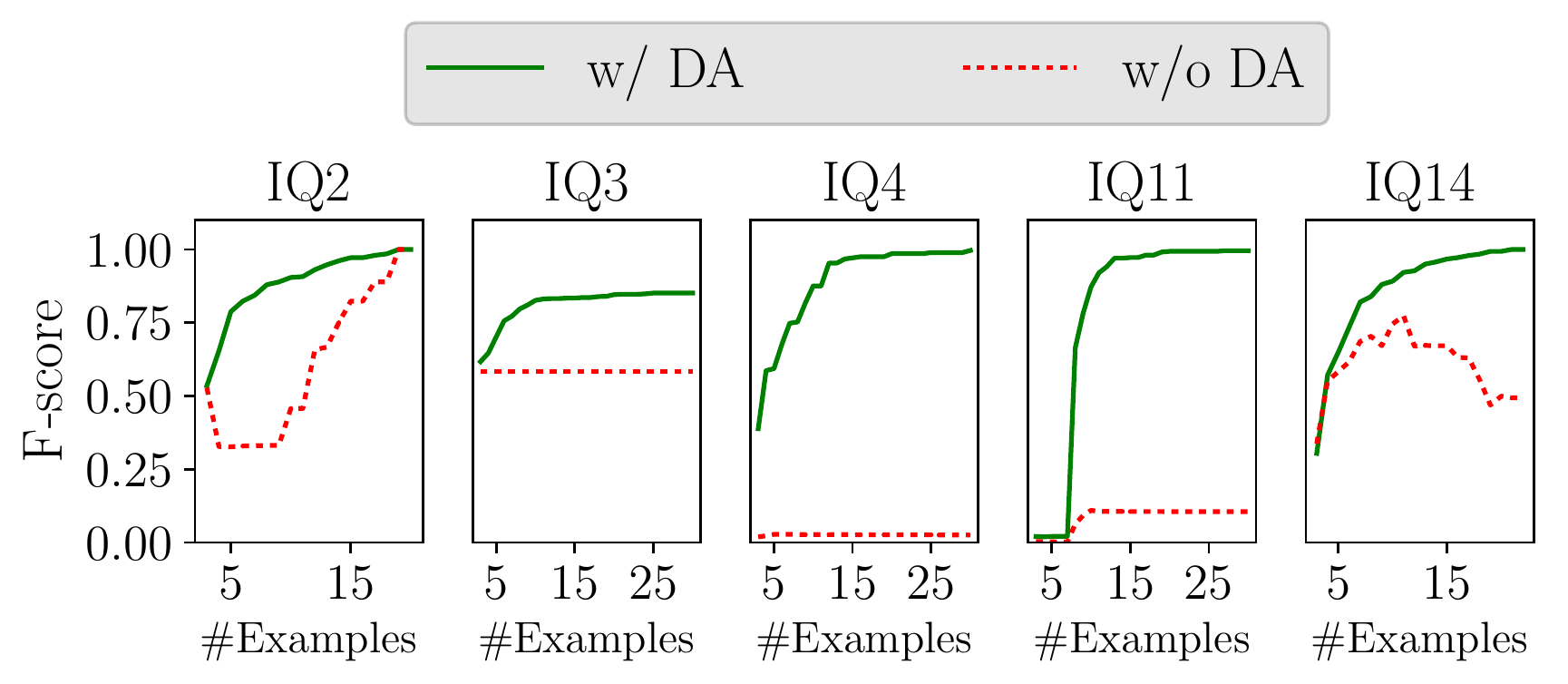}
	\vspace{-7mm}
    \caption{Effect of disambiguation on IMDb}
    \label{daEffect}
    \end{center}
    % \vspace{-2mm}
\end{figure}

\vspace{-2mm}
\paragraph*{Execution time}%\label{executionTimeQuality}
\looseness-1
While the accuracy results demonstrate that the abduced queries are
semantically close to the intended queries, \approach could be deriving a
query that is semantically close, but more complex and costly to compute. In
\figs~\ref{execTime} and~\ref{execTimedblp} we graph the average runtime of the
abduced queries and the actual benchmark queries. We observe that in most
cases the abduced queries and the corresponding benchmarks are similar in
execution time. Frequently, the abduced queries are faster because they take
advantage of the precomputed relations in the \adb. In few cases (\sql{IQ1},
\sql{IQ5}, and \sql{IQ7}) \approach discovered additional properties that,
while not specified by the original query, are inherent in all intended
entities. For example, in \sql{IQ5}, all movies with \sql{Tom Cruise} and
\sql{Nicole Kidman} are also English language movies and released between 1990
and 2014.

\vspace{-2mm}
\paragraph*{Effect of entity disambiguation} 
\looseness-1
Finally, we 
% evaluate the effect of entity disambiguation on abduction
% accuracy. We 
found that entity disambiguation never hurts abduction accuracy, and may
significantly improve it. \fig~\ref{daEffect} displays the impact of
disambiguation for five IMDb benchmark queries, where disambiguation
significantly improves the f-score.

\begin{figure}[t!]
    \begin{center}
    \includegraphics[trim={5mm 8mm 0mm 2mm},width=.48\textwidth]{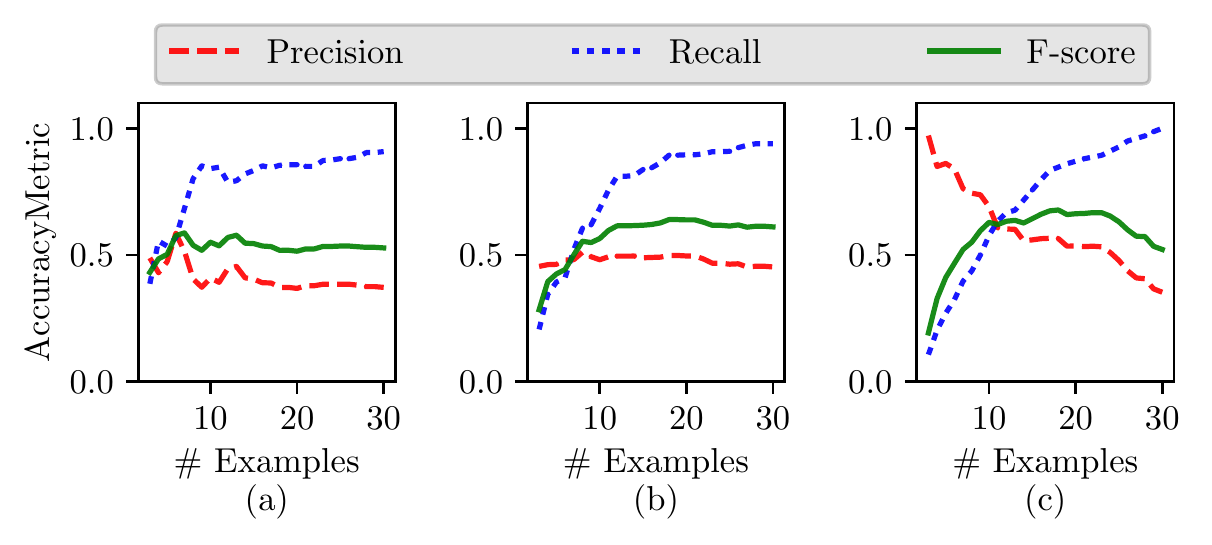}
	\vspace{-2mm}
    \caption{Precision, recall, and f-score for (a) Funny actors (b) 2000s Sci-Fi movies (c) Prolific DB researchers}
	\vspace{-6mm}
    \label{precisionRecallFscoreIMDbfunnyScifi}
    \end{center}
    \begin{center}
    \includegraphics[trim={5mm 6mm 0 0}, width=.48\textwidth]{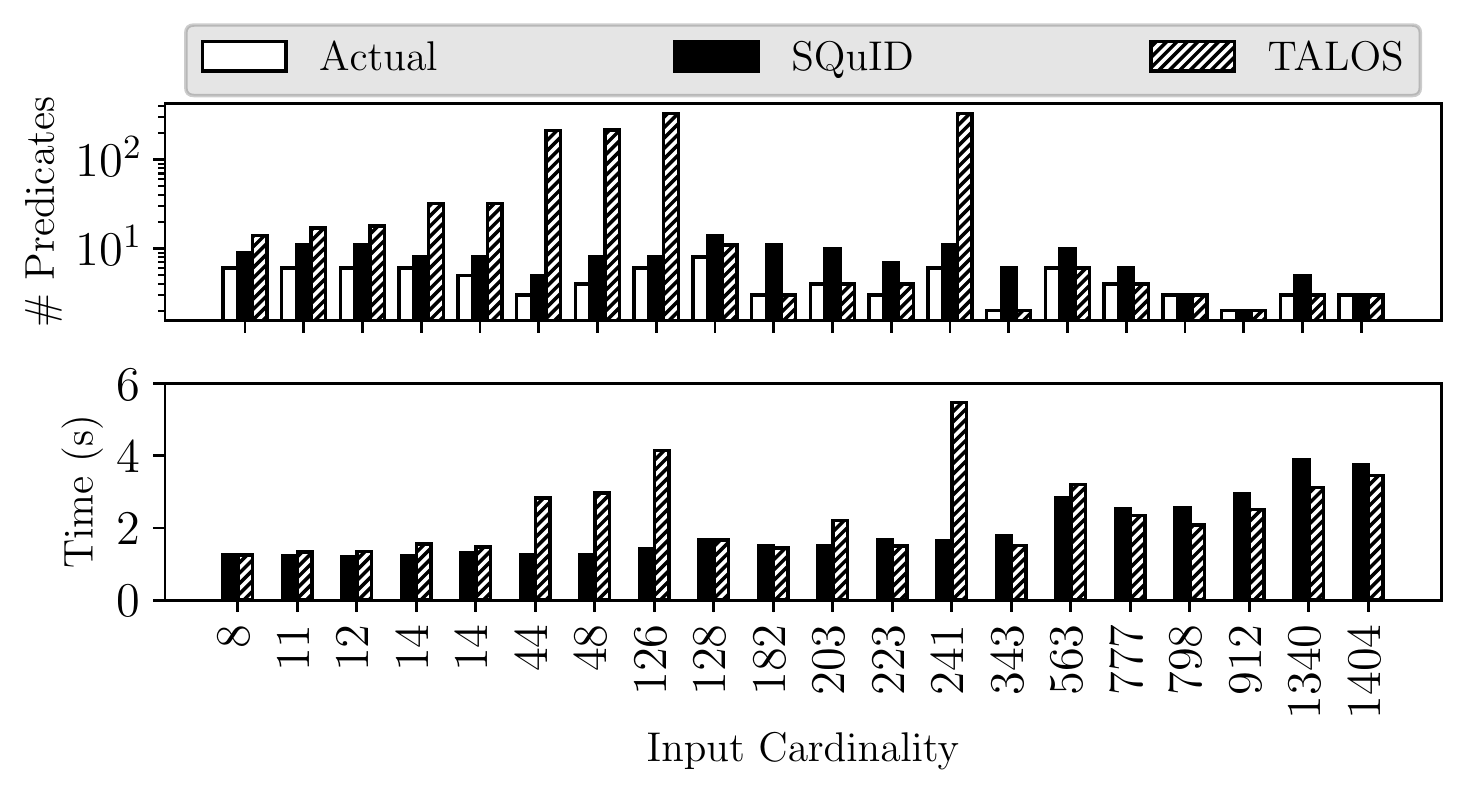}
    \vspace{-4mm}
    \caption{Both systems achieve perfect f-score on Adult (not shown).
    \approach produces significantly smaller queries, often by orders of magnitude, and is often much faster.
}
	\vspace{0mm}
    \label{talosComparisonPredCountAdult}
    \end{center}
\end{figure}

\subsection{Qualitative Case Studies}\label{qualitativeExp} 

In this section, we present qualitative results on the performance of
\approach, through a simulated user study.
We designed 3 case studies, by constructing queries and examples from
human-generated, publicly-available lists.

\noindent \textbf{Funny actors (IMDb).} We created a list of names of 211
``funny~actors'', collected from human-created public lists and Google
Knowledge Graph (sources are in \ifTechRep
Appendix~\ref{datasetAndBenchmark}\else \theTechRep \fi), and used these names
as examples of the query intent ``funny actors.''
\fig~\ref{precisionRecallFscoreIMDbfunnyScifi}(a) demonstrates the accuracy of
the abduced query over a varying number of
examples. Each data point is an
average across 10 different random samples of example sets of the corresponding
size. For this experiment, we tuned \approach to normalize the association
strength, which means that the relevant predicate would consider the fraction
of movies in an actor's portfolio classified as comedies, rather than the
absolute number.

\noindent
\textbf{2000s Sci-Fi movies (IMDb).} 
We used a user-created list of 165 Sci-Fi movies released in 2000s as examples
of the query intent ``2000s Sci-Fi movies''.
\fig~\ref{precisionRecallFscoreIMDbfunnyScifi}(b) displays the accuracy of the
abduced query, averaged across 10 runs for each example set size.

\noindent
\textbf{Prolific database researchers (DBLP).} 
We collected a list of data\-base researchers who served as chairs, group
leaders, or program committee members in SIGMOD 2011--2015 and selected the
top 30 most prolific. \fig~\ref{precisionRecallFscoreIMDbfunnyScifi}(c)
displays the accuracy of the abduced query averaged, across 10 runs for each
example set size.

\vspace{-2mm}
\paragraph*{Analysis} 

In our case studies there is no (reasonable) SQL query that models the intent
well and produces an output that exactly matches our lists. Public lists have
biases, such as not including less well-known entities even if these match the
intent.\footnote{ To counter this bias, our case study experiments use
popularity masks (derived from public lists) to filter the examples and the
abduced query outputs (\ifTechRep
Appendix~\ref{datasetAndBenchmark}\else\citeTechRep\fi). } In our prolific
researchers use case, some well-known and prolific researchers may happen to
not serve in service roles frequently, or their commitments may be in venues we
did not sample. Therefore, it is not possible to achieve high precision, as the
data is bound to contain and retrieve entities that don't appear on the lists,
even if the query is a good match for the intent. For this reason, our
precision numbers in the case studies are low. However our recall rises quickly
with enough examples, which indicates that the abduced queries converge to the
correct intent.

\begin{figure*}[t!]
    \centering
	\begin{subfigure}{.67\textwidth}
    \includegraphics[trim={6mm 6mm 1mm 0}, width=1\textwidth]{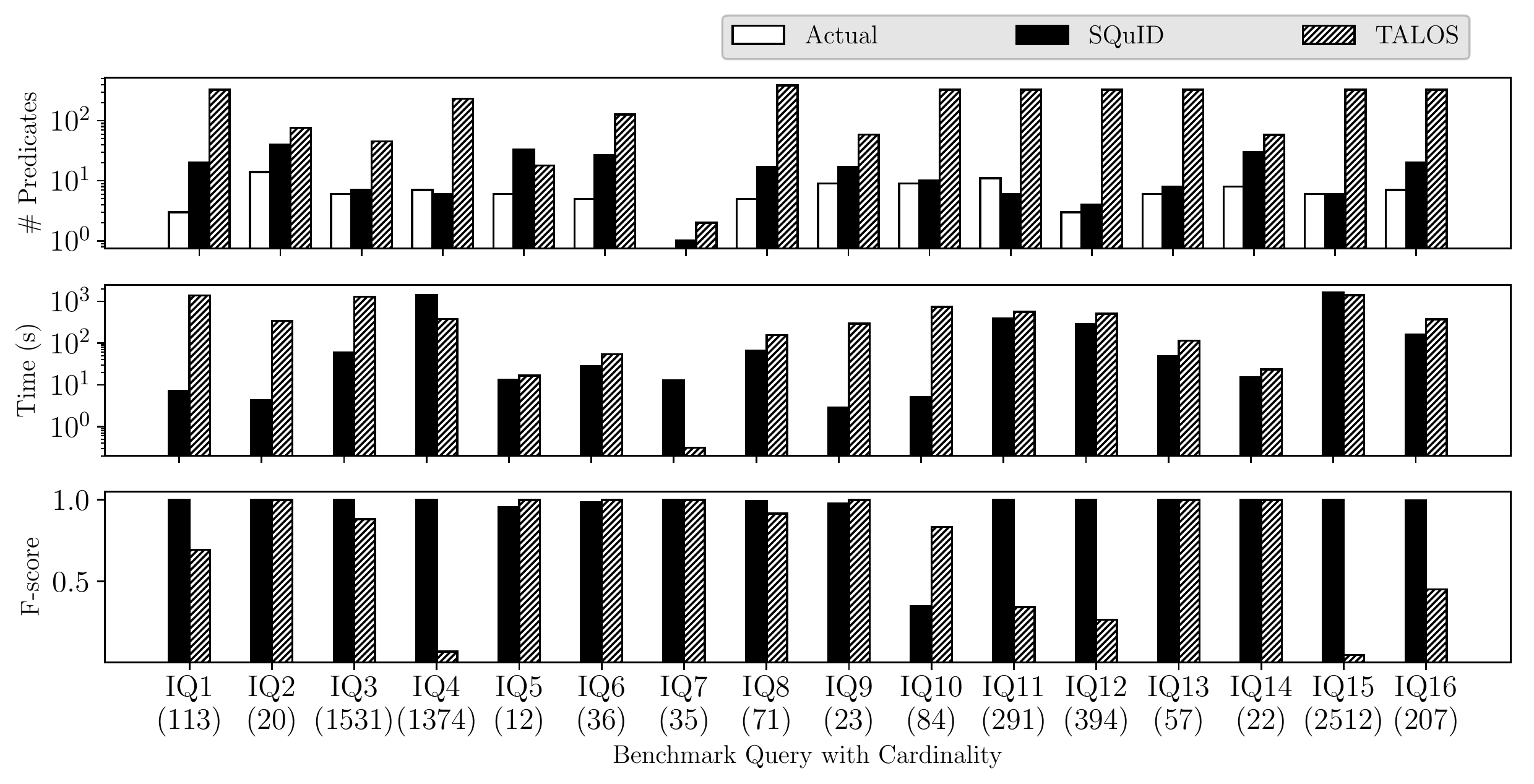}
    \caption{IMDb}
    \label{talosComparisonImdb}
	\end{subfigure}
	\vspace{-2mm}
	\begin{subfigure}{.25\textwidth}
    \includegraphics[trim={1mm 6mm 0 -10mm}, width=1\textwidth]{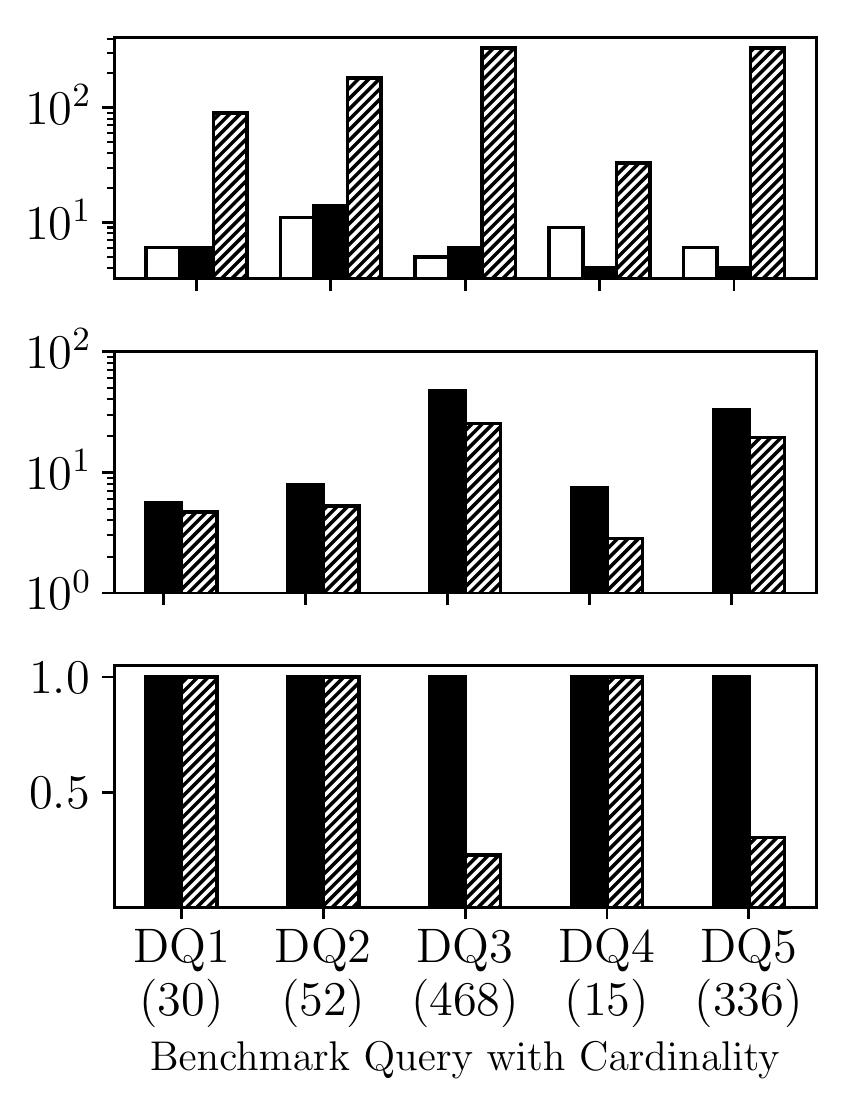}
    \caption{DBLP}
    \label{talosComparisonDBLP}
	\end{subfigure}
	\vspace{0mm}
	\caption{\approach produces queries with significantly fewer predicates than
TALOS and is more accurate on both IMDb and DBLP. \approach is almost always
faster on IMDb, but TALOS is faster on DBLP.
}
	\vspace{-2mm}
\end{figure*}

\subsection{Query Reverse Engineering}\label{qreCompareSection} 

We present an experimental comparison of \approach with
TALOS~\cite{TranVLDB2014}, a state-of-the-art Query Reverse Engineering (QRE)
system.\footnote{Other related methods either focus on more restricted
query classes~\cite{FastQRE, ZhangSIGMOD2013} or do not scale to data sizes
large enough for this evaluation~\cite{ZhangASE2013, WangPLDI2017}
(overview in Figure~\ref{relatedWorkMatrix}).}
QRE systems operate in a closed-world setting,
assuming that the provided examples comprise the entire query output.
In contrast, \approach assumes an open-world setting, and only needs a few
examples. In the closed-world setting, \approach is handicapped against a
dedicated QRE system, as it does not take advantage of the closed-world
constraint in its inference.

For this evaluation under the QRE setting, we use the IMDb and DBLP datasets,
as well as the Adult dataset, on which TALOS was shown to perform
well~\cite{TranVLDB2014}. For each dataset, we provided the entire output of
the benchmark queries as input to \approach and TALOS. Since there is no need
to drop coincidental filters for query reverse engineering, we set the
parameters so that \approach behaves optimistically (e.g., high filter prior,
low association strength threshold, etc.).\footnote{Details on the system
parameters are in \ifTechRep Appendix~\ref{parameterDiscussion}\else
\theTechRep \fi.} We adopt the notion of \emph{instance equivalent query} (IEQ)
from the QRE literature~\cite{TranVLDB2014} to express that two queries produce
the same set of results on a particular database instance. A QRE task is
successful if the system discovers an IEQ of the original query (f-score=1).
For the IMDb dataset, \approach was able to successfully reverse engineer 11
out of 16 benchmark queries. Additionally, in 4 cases where exact IEQs were not
abduced, \approach queries generated output with $\ge 0.98$ f-score. \approach
failed only for \sql{IQ10}, which is a query that falls outside the supported
query family, as discussed in Section~\ref{quality}. For the DBLP and Adult
datasets, \approach successfully reverse-engineered all benchmark queries.

\vspace{-2mm}
\paragraph*{Comparison with TALOS}

We compare \approach to TALOS on three metrics: number of predicates
(including join and selection predicates), query discovery time, and f-score.  

\noindent 
\textbf{Adult.}
Both \approach and TALOS achieved perfect f-score on the 20
benchmark queries. \fig~\ref{talosComparisonPredCountAdult} compares the
systems in terms of the number of predicates in the queries they produce (top)
and query discovery time (bottom). \approach almost always produces simpler
queries, close in the number of predicates to the original query, while TALOS
queries contain more than 100 predicates in 20\% of the cases.

\approach is faster than TALOS when the input cardinality is low ($\sim$100
tuples), and becomes slower for the largest input sizes ($> 700$ tuples).
\approach was not designed as a QRE system, and in
practice, users rarely provide large example sets. \approach's focus is on
inferring simple queries that model the intent, rather than cover all examples
with potentially complex and lengthy queries.

\noindent \textbf{IMDb.}
\fig~\ref{talosComparisonImdb} compares the two systems on the 16 benchmark
queries of the IMDb dataset. \approach produced better queries in almost all
cases: in all cases, our abduced queries where significantly smaller, and our
f-score is higher for most queries. \approach was also faster than TALOS for
most of the benchmark queries. We now delve deeper into some particular cases.

For \sql{IQ1} (cast of \sql{Pulp Fiction}), TALOS produces a query with f-score
= 0.7. We attempted to provide guidance to TALOS through a system parameter
that specifies which attributes to include in the selection predicates (which
would give it an unfair advantage). TALOS first performs a full join among the
participating relations (\sql{person} and \sql{castinfo}) and then performs
classification on the denormalized table (with attributes person, movie, role).
TALOS gives all rows referring to a cast member of \sql{Pulp Fiction} a
positive label (based on the examples), regardless of the movie that row refers
to, and then builds a decision tree based on these incorrect labels. This is a
limitation of TALOS, which \approach overcomes by looking at the semantic
similarities of the examples, rather than treating them simply as labels.

\looseness-1
\approach took more time than TALOS in \sql{IQ4}, \sql{IQ7}, and \sql{IQ15}.
The result sets of \sql{IQ4} and \sql{IQ15} are large ($>1000$), so this
is expected. \sql{IQ7}, which retrieves all movie genres, does not require any
selection predicate. As a decision tree approach, TALOS has the advantage
here, as it stops at the root and does not need to traverse the tree. In
contrast, \approach retrieves all semantic properties of the example tuples
only to discover that either there is nothing common among them, or the
property is not significant. While \approach takes longer, it
still abduces the correct query. These cases are not representative
of QBE scenarios, as users are unlikely to provide large number of example
tuples or have very general intents (PJ queries without selection).

\noindent \textbf{DBLP.} \fig~\ref{talosComparisonDBLP} compares the two
systems on the DBLP dataset. Here, \approach successfully reverse engineered
all five benchmark queries, but TALOS failed to reverse engineer two of them.
TALOS also produced very complex queries, with 100 or more predicates for four
of the cases. In contrast, \approach's abductions were orders of magnitude
smaller, on par with the original query. On this dataset, \approach was slower
than TALOS, but not by a lot.

\subsection{Comparison with learning methods}\label{comparePULearning}

Query intent discovery can be viewed as a one-class classification problem,
where the task is to identify the tuples that satisfy the desired intent.
Positive and Unlabeled (PU) learning addresses this problem setting by learning
a classifier from positive examples and unlabeled data in a semi-supervised
setting. We compare \approach against an established PU-learning
method~\cite{elkan2008learning} on 20 benchmark queries of the Adult dataset.
The setting of this experiment conforms with the technique's
requirements~\cite{elkan2008learning}: the dataset comprises of a single
relation and the examples are chosen uniformly at random from the positive data.

\begin{figure}[t!]
     \includegraphics[width=0.48\textwidth]{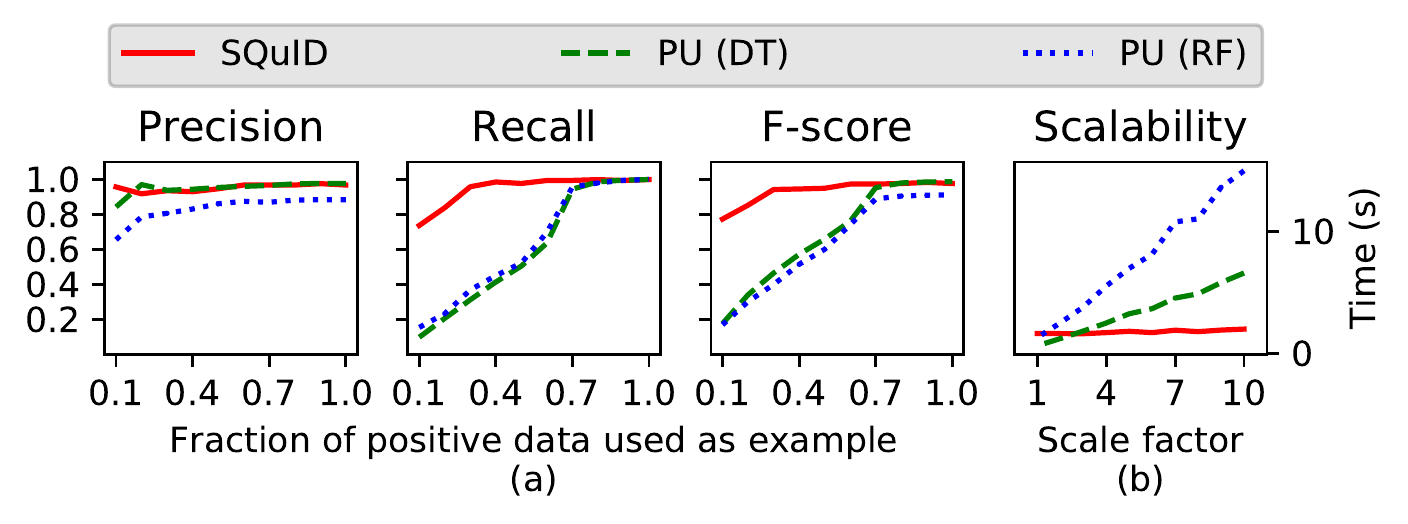} 
      \vspace{-8mm}
        \caption{(a) PU-learning needs a large fraction ($>70\%$) of the query
        results (positive data) as examples to achieve accuracy comparable to
        \approach. (b) The total required time for training and prediction in
        PU-learning increases linearly with the data size. In contrast,
        abduction time for \approach increases logarithmically.}
         \vspace{2mm}
         \label{mlComparison}   
\end{figure}

Figure~\ref{mlComparison}~(a) compares the accuracy of \approach and
PU-learning using two different estimators, decision tree (DT) and random
forest (RF). We observe that PU-learning needs a large fraction ($>70\%$) of
the query result to achieve f-score comparable to \approach. PU-learning favors
precision over recall, and the latter drops significantly when the number of
examples is low. In contrast, \approach achieves robust performance, even with
few examples, because it can encode problem-specific assumptions (e.g., that
there exists an underlying SQL query that models the intent, that some filters
are more likely than other filters, etc.); this cannot be done in
straightforward ways for machine learning methods.

To evaluate scalability, we replicated the Adult dataset, with a scale factor
up to 10x. Figure~\ref{mlComparison}~(b) shows that PU-learning becomes
significantly slower than \approach as the data size increases, whereas
\approach's runtime performance remains largely unchanged.
This is due to the fact that, \approach does not directly operate on the data
outside of the examples (unlabeled data); rather, it relies on the \adb, which
contains a highly compressed summary of the semantic property statistics (e.g.,
filter selectivities) of the data. In contrast, PU-learning builds a new
classifier over all of the data for each query intent discovery task. We
provide more discussion on the connections between \approach and machine
learning approaches in Section~\ref{relatedWork}.

\section{Related Work}\label{relatedWork}

\textbf{Query-by-Example (QBE)} was an early effort to assist users without SQL
expertise in formulating SQL queries~\cite{ZloofVLDB75}. Existing QBE
systems~\cite{ShenSIGMOD2014, PsallidasSIGMOD2015} identify relevant relations
and joins in situations where the user lacks schema understanding, but are
limited to project-join queries. These systems focus on the common structure of
the example tuples, and do not try to learn the common semantics as \approach
does. QPlain~\cite{DeutchICDE2016} uses user-provided provenance of the example
tuples to learn the join paths and improve intent inference. However, this
assumes that the user understands the schema, content, and domain to provide
these provenance explanations, which is often unrealistic for non-experts.

\textbf{Set expansion} is a problem corresponding to QBE in Knowledge
Graphs~\cite{ZhangSIGIR2017, DBLP:conf/icdm/WangC07, wordgrabbag}.
SPARQLByE~\cite{SPARQLByE2016}, built on top of a SPARQL QRE
system~\cite{SPARQLRE}, allows querying RDF datasets by annotated
(positive/negative) example tuples. In semantic knowledge graphs, systems
address the entity set expansion problem using maximal-aspect-based entity
model, semantic-feature-based graph query, and entity co-occurrence
information~\cite{LimEDBT2013, JayaramTKDE2015, HanICDE2016, MetzgerJIIS2017}.
These approaches exploit the semantic context of the example tuples, but they
cannot learn new semantic properties, such as aggregates involving numeric
values, that are not explicitly stored in the knowledge graph, and they cannot
express derived semantic properties without exploding the graph
size.\footnote{To represent ``appearing in more than K comedies'', the
knowledge graph would require one property for each possible value of $K$.}

\textbf{Interactive approaches} rely on relevance feedback on system-generated
tuples to improve query inference and result delivery~\cite{AbouziedPODS2013,
BonifatiTODS2016, DimitriadouTKDE2016, GeBigdata2016, LiVLDB2015}. Such systems
typically expect a large number of interactions, and are often not suitable for
non-experts who may not be sufficiently familiar with the data to provide
effective feedback.

\looseness-1 \textbf{Query Reverse Engineering (QRE)}~\cite{WeissPODS2017,
barcelICDT2017} is a special case of QBE that assumes that the provided
examples comprise the complete output of the intended query. Because of this
closed-world assumption, QRE systems can build data classification models on
denormalized tables~\cite{TranVLDB2014}, labeling the provided tuples as
positive examples and the rest as negative. Such methods are not suitable for
our setting, because we operate with few examples, under an open-world
assumption. While few QRE approaches~\cite{FastQRE} relax the closed world
assumption (known as the \emph{superset QRE} problem) they are also limited to
PJ queries similar to the existing QBE approaches. Most QRE methods are limited
to narrow classes of queries, such as PJ~\cite{ZhangSIGMOD2013, FastQRE},
aggregation without joins~\cite{TanVLDB2017}, or top-k
queries~\cite{PanevEDBT2016}. REGAL+\cite{regalPlus2018} handles SPJA queries
but only considers the schema of the example tuples to derive the joins and
ignores other semantics. In contrast, \approach considers joining relations
without attributes in the example schema (Example~\ref{collegeDBExample}).

A few QRE methods do target expressive SPJ queries~\cite{ZhangASE2013,
WangPLDI2017}, but they only work for very small databases ($< 100$ cells), and
do not scale to the datasets used in our evaluation. Moreover, the user needs
to specify the data in their entirety, thus expecting complete schema
knowledge, while SCYTHE~\cite{WangPLDI2017} also expects hints from the user
towards precise discovery of the constants of the query predicates.

\textbf{Machine learning} methods can model QBE settings as classification
problems, and relational machine learning targets relational settings in
particular~\cite{getoor2007introduction}. However, while the provided examples
serve as positive labels, QBE settings do not provide explicit negative
examples. Semi-supervised statistical relational learning
techniques~\cite{DBLP:conf/icdm/XiangN08} can learn from unlabeled and labeled
data, but require \emph{unbiased} sample of negative examples. There is no
straightforward way to obtain such a sample in our problem setting without
significant user effort.

Our problem setting is better handled by one-class
classification~\cite{manevitz2001one, khan2009survey}, more specifically,
Positive and Unlabeled (PU) learning~\cite{DBLP:conf/kdd/YuHC02,
liu2003building, DBLP:journals/corr/abs-1811-04820, elkan2008learning,
DBLP:conf/aaai/BekkerD18, mordelet2014bagging}, which learns from positive
examples and unlabeled data in a semi-supervised
setting~\cite{Chapelle:2010:SL:1841234}. Most PU-learning methods assume
denormalized data, but relational PU-leaning methods do exist. However, all
PU-learning methods rely on one or more strong
assumptions~\cite{DBLP:journals/corr/abs-1811-04820} (e.g., all unlabeled
entities are negative~\cite{DBLP:conf/acl/NeelakantanRM15}, examples are
selected completely at random~\cite{elkan2008learning,
DBLP:conf/ilp/BekkerD17}, positive and negative entities are naturally
separable~\cite{DBLP:conf/kdd/YuHC02, liu2003building, srinivasan2001aleph},
similar entities are likely from the same
class~\cite{DBLP:conf/aaai/KhotNS14}). These assumptions create a poor fit for
our problem setting where the example set is very small, it may exhibit user
biases, response should be real-time, and intents may involve deep semantic
similarity.

\textbf{Other approaches} that assist users in query formulation involve query
recommendation based on collaborative filtering~\cite{EirinakiTKDE2014}, query
autocompletion~\cite{KhoussainovaPVLDB2010}, and query
suggestion~\cite{FanICDE2011, YmalDBVLDB2013, JiangVLDB2015}. Another approach
to facilitating data exploration is keyword-based search~\cite{AgrawalICDE2002,
HristidisVLDB2002, ZengEDBT2016}. User-provided examples and interactions
appear in other problem settings, such as learning schema
mappings~\cite{SarmaICDT10, QianSIGMOD2012,BonifatiSIGMOD17}. The query
likelihood model in IR~\cite{DBLP:books/daglib/0021593} resembles our
technique, but does not exploit the similarity of the input entities.

\section{Summary and Future Directions}\label{discussion}

In this paper, we focused on the problem of query intent discovery from a set
of example tuples. We presented \approach, a system that performs query intent
discovery effectively and efficiently, even with few examples in most cases.
The insights of our work rely on exploiting the rich information present in
the data to discover similarities among the provided examples, and distinguish
between those that are coincidental and those that are intended. Our
contributions include a probabilistic abduction model and the design of an
abduction-ready database, which allow \approach to capture both explicit and
implicit semantic contexts. Our work includes an extensive experimental
evaluation of the effectiveness and efficiency of our framework over three
real-world datasets, case studies based on real user-generated examples and
abstract intents, and comparison with the state-of-the-art in query reverse
engineering (a special case of query intent discovery) and with PU-learning. Our
empirical results highlight the flexibility of our method, as it is extremely
effective in a broad range of scenarios. Notably, even though \approach
targets query intent discovery with a small set of a examples, it outperforms
the state-of-the-art in query reverse engineering in most cases, and is superior to learning techniques.

There are several possible improvements and research directions that can stem
from our work, including smarter semantic context inference using log data,
example recommendation to increase sample diversity and improve abduction,
techniques for adjusting the depth of association discovery, on-the-fly \adb
construction, and efficient \adb maintenance for dynamic datasets.

\balance

\nocite{dejong1979prediction, DBLP:conf/eccv/JiZLY12, becker1980semantic, vallet2006personalized}
	 
\bibliographystyle{abbrv}
\bibliography{squid} 

\ifTechRep
\begin{appendix}

\section{Domain selectivity impact}\label{domainSelectivitySignificance} We use
the notion of \emph{domain coverage} of a filter $\phi_{\prop{A, V, \theta}}$
to denote the fraction of values of $A$'s domain that $V$ covers. As an
example, for attribute \sql{age}, suppose that the domain consists of values
in the range $(0, 100]$, then the filter $\phi_{\prop{\mathtt{age},\mathtt{(40, 90]},\bot}}$ has
$50\%$ domain coverage and the filter $\phi_{\prop{\mathtt{age},\mathtt{(40, 45]},\bot}}$ has
$5\%$ domain coverage. We use a threshold $\eta > 0$ to specify how much domain
coverage does not reduce the domain selectivity impact $\delta$. After that
threshold, as domain coverage increases, $\delta$ decreases. We use another
parameter $\gamma \ge 0$ which states how strongly we want to penalize a filter
for having large domain coverage. The value of $\gamma = 0$ implies that we do
not penalize at all, i.e., all filters will have $\delta(\phi) = 1$. As
$\gamma$ increases, we reduce $\delta$ more for larger domain coverages. We
compute the domain selectivity impact using the equation below:
\begin{align*}
   \delta(\phi_{\prop{A, V, \theta}}) = \frac{1}{{\max(1, \frac{domainCoverage(V)}{\eta})}^\gamma}
\end{align*}

\section{Outlier impact}\label{outlierSignificance} Towards computing outlier
impact of a filter $\phi_{\prop{A, V, \theta}}$, we first compute skewness of
the association strength distribution $\Theta_A$ within the family of derived
filters involving attribute $A$; and then check whether $\theta$ is an outlier
among them. We compute sample skewness of $\Theta_A = (a_1, a_2, ..., a_n)$,
with sample mean $\bar{a}$ and sample standard deviation $s$, using the
standard formula:
\begin{align*}
skewness(\Theta_A) = \frac{n\sum_{i=1}^n(a_i - \bar{a})^3}{s^3 (n-1)(n-2)} 
\end{align*}

\begin{figure}[t]
	\centering
	\setlength\tabcolsep{1.5pt}
	{\footnotesize
	\begin{tabular}{|p{0.195\textwidth}|p{0.270\textwidth}|}
		\hline
		\rowcolor{vlightgray}
		\multicolumn{1}{|c|}{Title}&
		\multicolumn{1}{c|}{Source}\\
		\hline
		\hline
		IMDb dataset& \myurl{https://datasets.imdbws.com/}\\\hline
	    DBLP dataset& \myurl{https://data.mendeley.com/datasets/3p9w84t5mr}\\\hline
	    Adult dataset & \myurl{https://archive.ics.uci.edu/ml/datasets/adult}\\\hline
	    Physically strong actors & \myurl{https://www.imdb.com/list/ls050159844/}\\\hline
		Top 1000 Actors and Actresses\textsuperscript{*} & \myurl{http://www.imdb.com/list/ls058011111/}\\\hline
	    Sci-Fi Cinema in the 2000s & \myurl{http://www.imdb.com/list/ls000097375/}\\\hline
	    Funny Actors & \myurl{https://www.imdb.com/list/ls000025701/}\\\hline
	    100 Random Comedy Actors & \myurl{https://www.imdb.com/list/ls000791012/}\\\hline
	    BEST COMEDY ACTORS & \myurl{https://www.imdb.com/list/ls000076773/}\\\hline
	    115 funniest actors & \myurl{https://www.imdb.com/list/ls051583078/}\\\hline
	    Top 35 Male Comedy Actors & \myurl{https://www.imdb.com/list/ls006081748/}\\\hline
	    Top 25 Funniest Actors Alive & \myurl{https://www.imdb.com/list/ls056878567/}\\\hline
		the top funniest actors in hollywood today & \myurl{https://www.imdb.com/list/ls007041954/}\\\hline
		Google knowledge graph: Actors: Comedy & \myurl{https://www.google.com/search?q=funny+actors}\\\hline    
	    The Best Movies of All Time\textsuperscript{*} &   \myurl{https://www.ranker.com/crowdranked-list/the-best-movies-of-all-time}\\\hline
	    Top H-Index for Computer Science \& Electronics\textsuperscript{*}& \myurl{http://www.guide2research.com/scientists/}\\\hline 
		\multicolumn{2}{|r|} {\textsuperscript{*} Used as popularity mask} \\\hline  
\end{tabular}}
\vspace{-2mm}
\caption{Source of datasets}
\vspace{1mm}
\label{dataSource}
	\centering
	\setlength\tabcolsep{2pt} 

	\centering
    {\footnotesize
	\begin{tabular}{p{.03\textwidth}p{.12\textwidth}r || p{.03\textwidth}p{.12\textwidth}r} 	
		\thickhline
		\rowcolor{vlightgray}
		\multicolumn{6}{c}{\textbf{IMDb \& variations}}\\
		\hline		
			\rowcolor{vlightgray}
		\multicolumn{3}{c||}{IMDb} & \multicolumn{3}{c}{bd-IMDb}\\
		\hline
		& DB size & 633 MB &  &DB size & 1926 MB\\
		& \#Relations & 15 & & \#Relations & 15\\
		& Precomputed DB size & 2310 MB & &Precomputed DB size & 5971 MB\\
		& Precomputation time & 150 min & & Precomputation time & 370 min\\
		\hline
		& \sql{person} & 6,150,949 & & \sql{person} & 12,301,898\\
		Rel. & \sql{movie} & 976,719 & Rel. & \sql{movie} & 1,953,438\\
		Card. & \sql{castinfo} & 14,915,325 & Card. & \sql{castinfo} & 59,661,300\\
		\hline
		\hline
			\rowcolor{vlightgray}
 		\multicolumn{3}{c||}{bs-IMDb} & \multicolumn{3}{c}{sm-IMDb}\\
 		\hline
 		& DB size & 1330 MB & & DB size & 75 MB\\
		& \#Relations & 15 & & \#Relations & 15\\
 		& Precomputed DB size & 4831 MB & & Precomputed DB size  & 317 MB\\
 		& Precomputation time & 351 min & & Precomputation time & 14 min\\
 		\hline
 		& \sql{person} & 12,301,898 & & \sql{person} & 65,865\\
 		Rel. & \sql{movie} & 1,953,438 & Rel. & \sql{movie} & 335,705\\
 		Card. & \sql{castinfo} & 29,830,650 & Card. & \sql{castinfo} & 1,364,890\\
		\thickhline
			\rowcolor{vlightgray}
		\multicolumn{3}{c||}{\textbf{DBLP}} & \multicolumn{3}{c}{\textbf{Adult}}\\
		\thickhline
		& DB size & 22 MB & & DB size & 4 MB\\
		& \#Relations & 14 & & \#Relations & 1\\
		& Precomputed DB size & 98 MB & & Precomputed DB size  & 5 MB\\
		& Precomputation time & 42 min & & Precomputation time & 3 min\\ 
		\hline
		& \sql{author} & 126,094 & & &\\
		Rel. & \sql{publication} & 148,521 & Rel. & \sql{adult} & 32,561\\
		Card. & \sql{authortopub} & 416,445 & Card. & &\\
		\thickhline
	\end{tabular}} \\ 
	\vspace{-1mm}
	\caption{Description of IMDb, DBLP, and Adult datasets}
	\vspace{2mm}
	\label{datasetDescription}
\end{figure}

A distribution is skewed if its skewness exceeds a threshold $\tau_s$. For
outlier detection, we use the mean and standard deviation method. For sample
mean $\bar{a}$, sample standard deviation $s$, and a constant $k \ge 2$, $a_i$
is an outlier if ($a_i - \bar{a}) > ks$. For $n < 3$, skewness is not defined
and we assume all elements to be outliers. We compute outlier impact
$\lambda(\phi_{\prop{A, V, \theta}})$:
\begin{align*}
	\lambda(\phi_{\prop{A, V, \theta}}) = \begin{cases}
										1	\;\;\; \text{ $\theta{ = }\bot \vee (skewness(\Theta_A){ > }\tau_s \wedge outlier(\theta))$}\\
										0	\;\;\; \text{ otherwise}
									\end{cases}
\end{align*}

\section{Proof of Theorem~1} \label{theoremProof}
\begin{proof}[Proof] 
We will prove Theorem~\ref{theTheorem} by contradiction. Suppose that
$\FilterSubset$ is the optimal set of filters, i.e., $Q^\FilterSubset$ is the
most likely query. Additionally, suppose that $\FilterSubset$ is the minimal
set of filters for obtaining such optimality, i.e., $\nexists
\FilterSubset^\prime$ such that $|\FilterSubset^\prime| < \FilterSubset \wedge
Pr(Q^{\FilterSubset^\prime}|E) = Pr(Q^\FilterSubset|E)$. Now suppose that,
Algorithm~\ref{squidAlgo} returns a sub-optimal query
$Q^{\FilterSubset^\prime}$, i.e., $Pr(Q^{\FilterSubset^\prime}|E) <
Pr(Q^\FilterSubset|E)$. Since $Q^{\FilterSubset^\prime}$ is suboptimal,
$\FilterSubset^\prime \neq \FilterSubset$; therefore at least one of the
following two cases must hold:

\noindent \textbf{Case 1:} $\exists\phi$ such that $\phi\in\FilterSubset \wedge
\phi \not\in \FilterSubset^\prime$. Since Algorithm~\ref{squidAlgo} did not
include $\phi$, it must be the case that $include_\phi \le exclude_\phi$.
Therefore, we can exclude $\phi$ from $\FilterSubset$ to obtain $\FilterSubset
- \{\phi\}$ and according to \eqn~\ref{mainDerivation}, $Pr(Q^{\FilterSubset-
\{\phi\}}|E) \ge Pr(Q^{\FilterSubset}|E)$ which contradicts with our assumption
about the optimality and minimality of $\FilterSubset$.

\noindent \textbf{Case 2:} $\exists\phi$ such that $\phi\not\in\FilterSubset
\wedge \phi \in \FilterSubset^\prime$. Since Algorithm~\ref{squidAlgo} included
$\phi$, it must be the case that $include_\phi > exclude_\phi$. Therefore, we
can add $\phi$ to $\FilterSubset$ to obtain $\FilterSubset \cup \{\phi\}$ and
according to \eqn~\ref{mainDerivation}, $Pr(Q^{\FilterSubset \cup \{\phi\}}|E)
> Pr(Q^{\FilterSubset}|E)$ which again contradicts with our assumption about
the optimality of $\FilterSubset$.

Hence, $Q^{\FilterSubset^\prime}$ cannot be suboptimal and this implies that
Algorithm~\ref{squidAlgo} returns the most likely query. 
\end{proof}
Note that, in a special case where $include_\phi = exclude_\phi$,
Algorithm~\ref{squidAlgo} drops the filter using Occam's razor principle to
keep the query as simple as possible. But this, however, does not return any
query that is strictly less likely than the best query.

\section{Datasets and Benchmark Queries}\label{datasetAndBenchmark} 

We collect the datasets from various sources and provide them in
Figure~\ref{dataSource}. The detailed description of the datasets are given in
\fig~\ref{datasetDescription}. We mention the cardinalities of the big
relations for providing a sense of the data and their associations.

\subsection{Alternative IMDb Datasets} For the scalability experiment, we
generated 3 versions of the IMDb database. For obtaining a downsized database
sm-IMDb, we dropped persons with less than 2 affiliated movies and/or who have
too many semantic information missing, and movies that have no cast
information. We produced two upsized databases: one with dense associations
bd-IMDb, and the other with sparse associations bs-IMDb. bd-IMDb contains
duplicate entries for all movies, persons, and companies (with different
primary keys), and the associations among persons and movies are duplicated to
produce more dense associations. For example, if \sql{P1} acted in
\sql{M1} in IMDb, i.e., \sql{(P1,M1)} exists in IMDB's \sql{castinfo},
we added a duplicate person \sql{P2}, a duplicate movie \sql{M2}, and 3
new associations, \sql{(P1,M2)}, \sql{(P2,M2)}, and \sql{(P2,M1)}, to
bd-IMDb's \sql{cast\-info}. For bs-IMDb, we only duplicated the old
associations, i.e., we added \sql{P2} and \sql{M2} in a similar fashion,
but only added \sql{(P2,M2)} in \sql{castinfo}.

\subsection{Benchmark Queries}\label{benchmarkQueries} We discuss the benchmark
queries for all datasets here. \figs~\ref{imdbBenchmarks}
and~\ref{dblpBenchmarks} display benchmark queries that we use to run different
experiments on the IMDb and DBLP datasets, respectively. The tables show the
query intents, details of the corresponding queries in SQL (number of joining
relations (J) and selection predicates (S)), and the result set cardinality.
\fig~\ref{adultBenchmarks} shows 20 benchmark queries along with their result
set cardinality for the Adult dataset.

\section{{\large\textbf{\approach}} Parameters}\label{parameterDiscussion} We
list the four most important \approach parameters in
\fig~\ref{systemParameters} along with brief description. We now discuss the
impact of these parameters on \approach and provide few empirical results.

\smallskip

\indent $\bm{\rho}$. The base filter prior parameter $\rho$ defines
\approach's tendency towards including filters. Small $\rho$ makes \approach
pessimistic about including a filter, and thus favors recall. In contrast,
large $\rho$ makes \approach optimistic about including a filter, which favors
precision. Low $\rho$ helps in getting rid of coincidental filters,
particularly with very few example tuples. However, with sufficient example
tuples, coincidental filters eventually disappears, and the effect of $\rho$
diminishes. \fig~\ref{rhoEffect} shows effect of varying the value of $\rho$
for few benchmark queries on the IMDb dataset. While low $\rho$ favors some
queries (\sql{IQ2}, \sql{IQ16}), it causes accuracy degradation for some
other queries (\sql{IQ3}, \sql{IQ4}, \sql{IQ11}), where high $\rho$
works better. It is a tradeoff and we found empirically that moderate value of
$\rho$ (e.g., 0.1) works best on an average.

\looseness-1
\indent $\bm{\gamma}$. The domain coverage penalty parameter $\gamma$
specifies \approach's leniency towards filters with large domain coverage. Low
$\gamma$ penalizes filters with large domain coverage less, and high $\gamma$
penalizes them more. \fig~\ref{gammaEffect} shows the effect of varying
$\gamma$. Very low $\gamma$ favors some queries (\sql{IQ3}, \sql{IQ4},
\sql{IQ11}) but also causes accuracy degradation for some other queries
(\sql{IQ2}, \sql{IQ16}), where high $\gamma$ works better. Like $\rho$,
it is also a tradeoff, and empirically we found moderate values of $\gamma$
(e.g., 2) to work well on an average.

\indent $\bm{\tau_a}$. The association strength threshold $\tau_a$ is
required to define the association strength impact $\alpha(\phi)$
(Section~\ref{queryPriorSection}). \fig~\ref{imdbVaryATIQ1} illustrates the effect of
different values of $\tau_a$ on the benchmark query \sql{IQ5} on the IMDb
dataset. The figure shows that, with very few example tuples, high $\tau_a$ is
preferable, since it helps dropping coincidental filters with weak
associations. Similar to other parameters, with increased number of example
tuples, the effect of $\tau_a$ diminishes.

\smallskip

\indent $\bm{\tau_s}$. The skewness threshold $\tau_s$ is required to
classify an association strength distribution as skewed or not
(Appendix~\ref{outlierSignificance}). \fig~\ref{imdbVarySKIQ1} illustrates the
effect of different values of $\tau_s$ on the benchmark query \sql{IQ1} on
the IMDb dataset. $\tau_s = N/A$ refers to the experiment where outlier impact
was not taken into account (i.e., $\lambda(\phi) = 1$ for all filters). In this
query, there were a number of unintended derived filters involving
\sql{certificate} and high $\tau_s$ helped to get rid of those. We also
found high $\tau_s$ to be very useful when we cannot use high $\tau_a$ due to
the nature of the query intent (e.g., \sql{IQ3}). However, too high $\tau_s$
is also not desirable, since it will underestimate some moderately skewed
distributions and drop intended filters. Empirically, we found that moderate
$\tau_s$ (e.g., 2--4) to work well on an average.

\begin{figure}[t]
	\centering
	\setlength\tabcolsep{1.5pt}
	{\footnotesize
	\begin{tabular}{|l|p{.33\textwidth}|c|c|r|}
		\hline
			\rowcolor{vlightgray}
		\multicolumn{1}{|c|}{ID}&
		\multicolumn{1}{c|}{Task}&
		\multicolumn{1}{c|}{J}& 
		\multicolumn{1}{c|}{S}& 
		\multicolumn{1}{c|}{\#Result}\\
		\hline
		\hline
		IQ1 & Entire cast of Pulp Fiction & 3 & 1 & 113 \\\hline
		IQ2 & Actors who appeared in all of The Lord of the Rings trilogy & 8 & 7 & 20 \\\hline
		IQ3 & Canadian actresses born after 1970 & 3 & 4 & 1531 \\\hline		
		IQ4 & Sci-Fi movies released in USA in 2016 & 5 & 3 & 1374 \\\hline		
		IQ5 & Movies Tom Cruise and Nicole Kidman acted together & 5 & 2 & 12 \\\hline 
		IQ6 & Movies directed by Clint Eastwood & 4 & 2 & 36\\\hline 
		IQ7 & All movie genres & 1 & 0 & 35 \\\hline 
		IQ8 & Movies by Al Pacino & 4 & 2 & 71 \\\hline 
		IQ9\textsuperscript{*} & Indian actors who acted in at least 15 Hollywood movies & 6 & 4 & 23 \\\hline 
		IQ10\textsuperscript{*} & Actors who acted in more than 10 Russian movies after 2010 & 6 & 4 & 84\\\hline 
		IQ11 & Hollywood Horror-Drama movies in 2005 -- 2008 & 7 & 5 & 291 \\\hline 
		IQ12 & Movies produced by Walt Disney Pictures & 3 & 1 &  394 \\\hline 
		IQ13 & Animation movies produced by Pixar & 5 & 2 & 57 \\\hline 
		IQ14 & Sci-Fi movies acted by Patrick Stewart& 6 & 3 & 22 \\\hline 
		IQ15 & Japanese Animation movies & 5 & 2 & 2512 \\\hline 
		IQ16\textsuperscript{*} & Walt Disney Pictures movies with more than 15 American cast members & 5 & 3 & 207 \\\hline 		           
		\multicolumn{5}{|r|} { {* Includes \sql{GROUP BY} and \sql{HAVING} clauses}} \\\hline
\end{tabular}}
\vspace{-2mm}
\caption{Benchmark queries for the IMDb dataset}
\vspace{1mm}
\label{imdbBenchmarks}
	\centering
	\setlength\tabcolsep{2pt} 
	\centering
	{\footnotesize
	\begin{tabular}{|l|p{.32\textwidth}|c|c|r|} 
		\hline
			\rowcolor{vlightgray}
		\multicolumn{1}{|c|}{ID}&
		\multicolumn{1}{c|}{Task}&
		\multicolumn{1}{c|}{J}& 
		\multicolumn{1}{c|}{S}& 
		\multicolumn{1}{c|}{\#Result}\\
		\hline
		\hline
		DQ1 & Authors who collaborated with both U Washington and Microsoft Research Redmond & 5 & 2 & 30\\\hline
		DQ2\textsuperscript{*} & Authors with at least 10 SIGMOD and at least 10 VLDB publications & 8 & 4 & 52\\\hline
		DQ3 & SIGMOD publications in 2010 -- 2012 & 3 & 3 & 468\\\hline		
		DQ4 & Publications Jiawei Han, Xifeng Yan, and Philip S. Yu published together & 7 & 3 & 15\\\hline		
		DQ5 & Publications between USA and Canada & 5 & 2 & 336\\\hline
		\multicolumn{5}{|r|} { {\textsuperscript{*} Includes \sql{GROUP BY}, \sql{HAVING},  and \sql{INTERSECT}}} \\\hline		
\end{tabular}}
\vspace{-2mm}
\caption{Benchmark queries for the DBLP dataset}
\vspace{1mm}
\label{dblpBenchmarks}
	
	\centering
	{\small
	\begin{tabular}{ccl} 
		\thickhline
			\rowcolor{vlightgray}
		\multicolumn{1}{c}{Para\-meter}&\multicolumn{1}{c}{Default value}&\multicolumn{1}{c}{Description}\\		
		\hline
		$\rho$ 		& 0.1			& Base filter prior parameter.\\
		$\gamma$ 	& 2				& Domain coverage penalty parameter.\\
		$\tau_a$ 	& 5				& Association strength threshold.\\		
		$\tau_s$ 	& 2.0			& Skewness threshold.\\		
	\thickhline
	\end{tabular}
	}
	\caption{List of \approach parameters with description.}
	\label{systemParameters}
	\vspace{2mm}
\end{figure}

\section{Analysis of prior art} 
\label{app:matrix}

We provided a summary of prior work to contrast with \approach in the
comparison matrix of Figure~\ref{relatedWorkMatrix}. In this section we explain
the comparison metrics and highlight the key differences among different
classes of query by example techniques and their variants.

We organize the prior work into three categories --- QBE (query by example),
QRE (query reverse engineering), and DX (data exploration). Furthermore, we
group QBE methods into two sub-categories, methods for relational databases,
and methods for knowledge graphs. All QRE and DX methods that we discuss are
developed on relational databases.

Finally, we provide an extensive discussion to contrast \approach against
existing semi-supervised machine learning approaches.

\subsection{Comparison Metrics} 
\textbf{Query class} encodes the expressivity of a query. We use four primitive
SQL operators (join, projection, selection, and aggregation) as comparison
metrics. Although all of these operators are not directly supported by data
retrieval mechanisms (e.g., graph query, SPARQL) for knowledge graphs, they
support similar expressivity through alternative equivalent operators.

\setlength\tabcolsep{2pt} 
\begin{figure}[H]
    {\footnotesize
	\begin{tabular}{|p{.42\textwidth}|r|} 
		\hline
			\rowcolor{vlightgray}
		\multicolumn{1}{|c}{SQL Query} &		\multicolumn{1}{|c|}{\#Result}\\
		\hline
		\hline
\footnotesize\sql{SELECT DISTINCT name FROM adult WHERE education = `Bachelors' AND occupation = `Craft-repair' AND hoursperweek >= 36 AND hoursperweek <= 40 AND age >= 46 AND age <= 47} & 8\\
\hline
\footnotesize\sql{SELECT DISTINCT name FROM adult WHERE race = `White' AND sex = `Female' AND nativecountry = `United-States' AND relationship = `Other-relative' AND occupation = `Machine-op-inspct' AND workclass = `Private'} & 11\\
\hline
\footnotesize\sql{SELECT DISTINCT name FROM adult WHERE occupation = `Craft-repair' AND workclass = `Private' AND age >= 65 AND age <= 68 AND relationship = `Husband'} & 12\\
\hline
\footnotesize\sql{SELECT DISTINCT name FROM adult WHERE maritalstatus = `Divorced' AND capitalgain >= 7298 AND capitalgain <= 10520 AND hoursperweek >= 40 AND hoursperweek <= 44 AND relationship = `Not-in-family'} & 14\\
\hline
\footnotesize\sql{SELECT DISTINCT name FROM adult WHERE capitalgain >= 4101 AND capitalgain <= 4650 AND workclass = `Private' AND age >= 41 AND age <= 44} & 14\\
\hline
\footnotesize\sql{SELECT DISTINCT name FROM adult WHERE occupation = `Protective-serv' AND hoursperweek >= 45 AND hoursperweek <= 48} & 44\\
\hline
\footnotesize\sql{SELECT DISTINCT name FROM adult WHERE education = `10th' AND race = `White' AND fnlwgt >= 334113 AND fnlwgt <= 403468} & 48\\
\hline
\footnotesize\sql{SELECT DISTINCT name FROM adult WHERE nativecountry = `United-States' AND hoursperweek >= 43 AND hoursperweek <= 45 AND race = `White' AND fnlwgt >= 106541 AND fnlwgt <= 118876} & 126\\
\hline
\footnotesize\sql{SELECT DISTINCT name FROM adult WHERE race = `White' AND education = `Bachelors' AND nativecountry = `United-States' AND capitalgain >= 6097 AND capitalgain <= 7688 AND maritalstatus = `Married-civ-spouse' AND relationship = `Husband'} & 128\\
\hline
\footnotesize\sql{SELECT DISTINCT name FROM adult WHERE education = `Bachelors' AND capitalloss >= 1848 AND capitalloss <= 1980} & 182\\
\hline
\footnotesize\sql{SELECT DISTINCT name FROM adult WHERE sex = `Male' AND nativecountry = `United-States' AND capitalloss >= 1848 AND capitalloss <= 1887} & 203\\
\hline
\footnotesize\sql{SELECT DISTINCT name FROM adult WHERE education = `Doctorate' AND maritalstatus = `Married-civ-spouse' AND nativecountry = `United-States'} & 223\\
\hline
\footnotesize\sql{SELECT DISTINCT name FROM adult WHERE education = `HS-grad' AND workclass = `Private' AND hoursperweek >= 45 AND hoursperweek <= 46 AND relationship = `Husband'} & 241\\
\hline
\footnotesize\sql{SELECT DISTINCT name FROM adult WHERE capitalgain >= 7688 AND capitalgain <= 8614} & 343\\
\hline
\footnotesize\sql{SELECT DISTINCT name FROM adult WHERE education = `Bachelors' AND maritalstatus = `Never-married' AND workclass = `Private' AND hoursperweek >= 40 AND hoursperweek <= 43 AND race = `White'} & 563\\
\hline
\footnotesize\sql{SELECT DISTINCT name FROM adult WHERE education = `HS-grad' AND nativecountry = `United-States' AND occupation = `Machine-op-inspct' AND race = `White'} & 777\\
\hline
\footnotesize\sql{SELECT DISTINCT name FROM adult WHERE nativecountry = `United-States' AND age >= 60 AND age <= 62} & 798\\
\hline
\footnotesize\sql{SELECT DISTINCT name FROM adult WHERE fnlwgt >= 271962 AND fnlwgt <= 288781} & 912\\
\hline
\footnotesize\sql{SELECT DISTINCT name FROM adult WHERE maritalstatus = `Married-civ-spouse' AND fnlwgt >= 221366 AND fnlwgt <= 259301} & 1340\\
\hline
\footnotesize\sql{SELECT DISTINCT name FROM adult WHERE maritalstatus = `Never-married' AND fnlwgt >= 185624 AND fnlwgt <= 211177} & 1404\\
\hline
\end{tabular}}
\vspace{-2mm}
\caption{Benchmark queries for the Adult dataset}
\vspace{-2.5mm}
\label{adultBenchmarks}
\end{figure}

\textbf{Semi-join} is a special type of join which is particularly useful for
QBE systems. A system is considered to support semi-join if it allows inclusion
of relations in the output query that have no attribute projected in the input
schema (e.g., in Example~\ref{collegeDBExample}, no attribute of
\sql{research} appears in the input tuples, but \sql{Q2} includes
\sql{research}). While knowledge graph based systems do not directly support
semi-join as defined in the relational database setting, they support same
expressivity through alternative mechanism.

\textbf{Implicit property} refers to the the properties that are not directly
stated in the data (e.g., number of comedies an actor appears in). In
\approach, we compute implicit properties by aggregating direct properties of
affiliated entities.

\begin{figure}[t!] 
   
    \includegraphics[width=.47\textwidth]{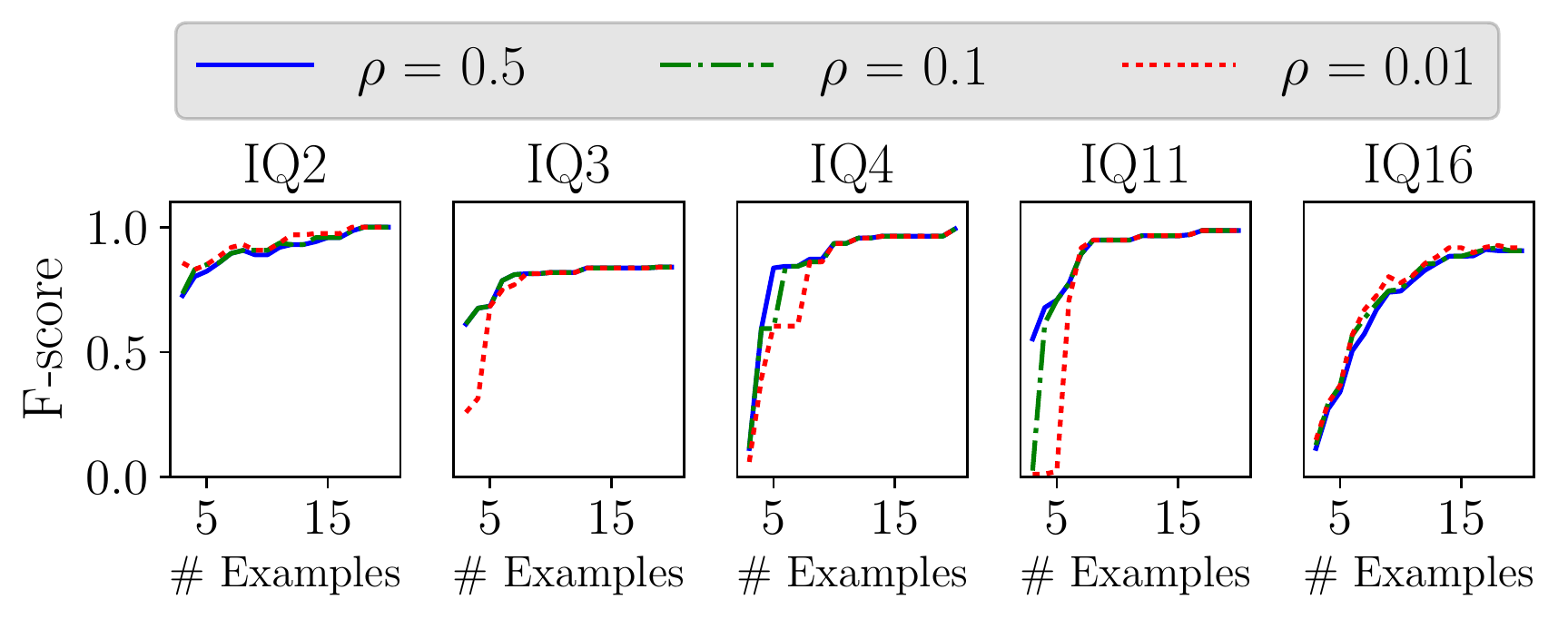}
    \caption{Effect of different values for $\rho$ for few benchmark queries of the IMDb dataset}
	\vspace{2mm}
    \label{rhoEffect}

    \includegraphics[trim={2mm 0mm 0 0mm},width=.47\textwidth]{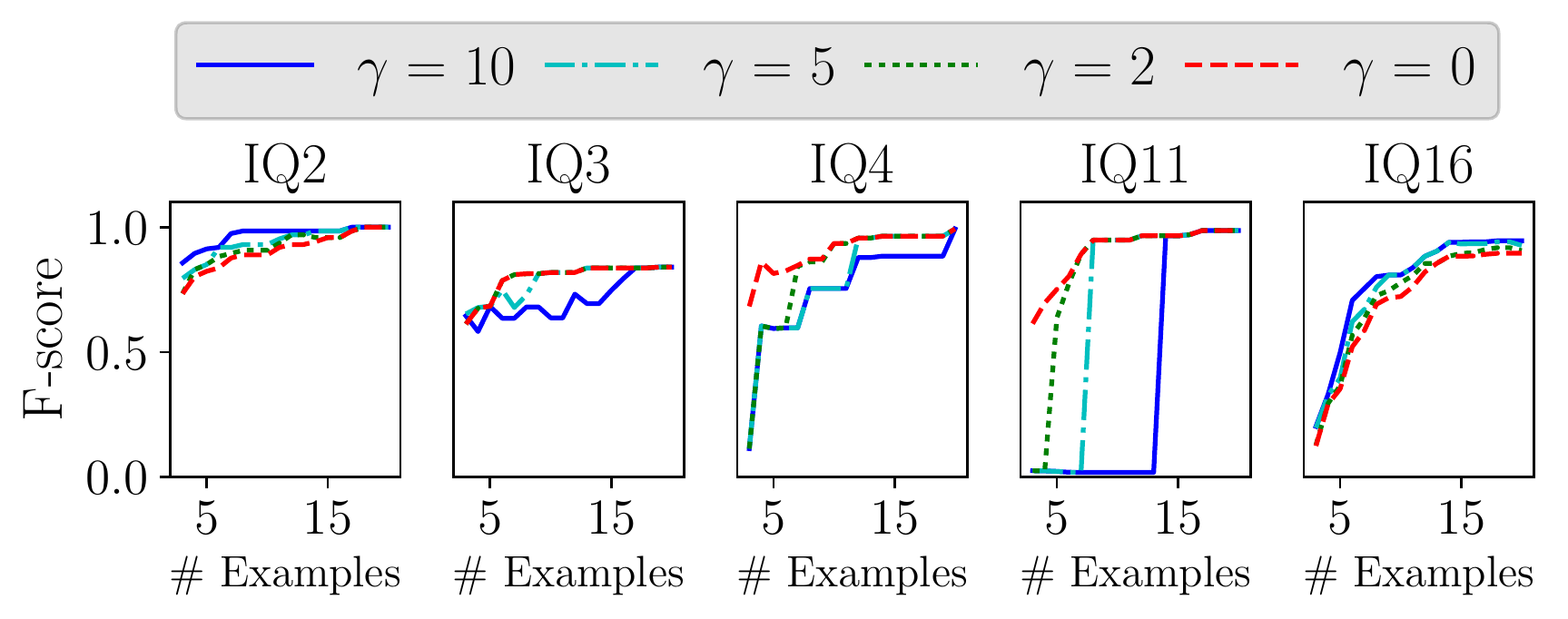}
    \caption{Effect of different values for $\gamma$ for few benchmark queries of the IMDb dataset}
	\vspace{2mm}
    \label{gammaEffect}

    \includegraphics[trim={2mm 0mm 0 0mm}, width=.47\textwidth]{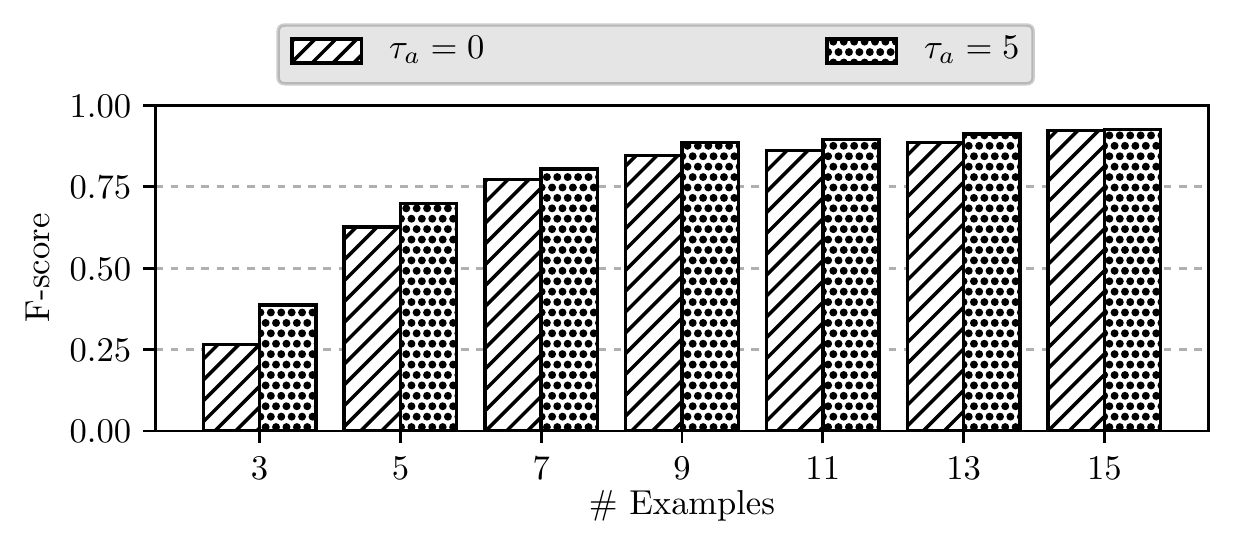}
	\vspace{-2mm}
  	\caption{Effect of different values for $\tau_a$ for \sql{IQ5} (IMDb)}
  	\vspace{2mm}
  	\label{imdbVaryATIQ1}

    \includegraphics[trim={2mm 0mm 0 0mm}, width=.47\textwidth]{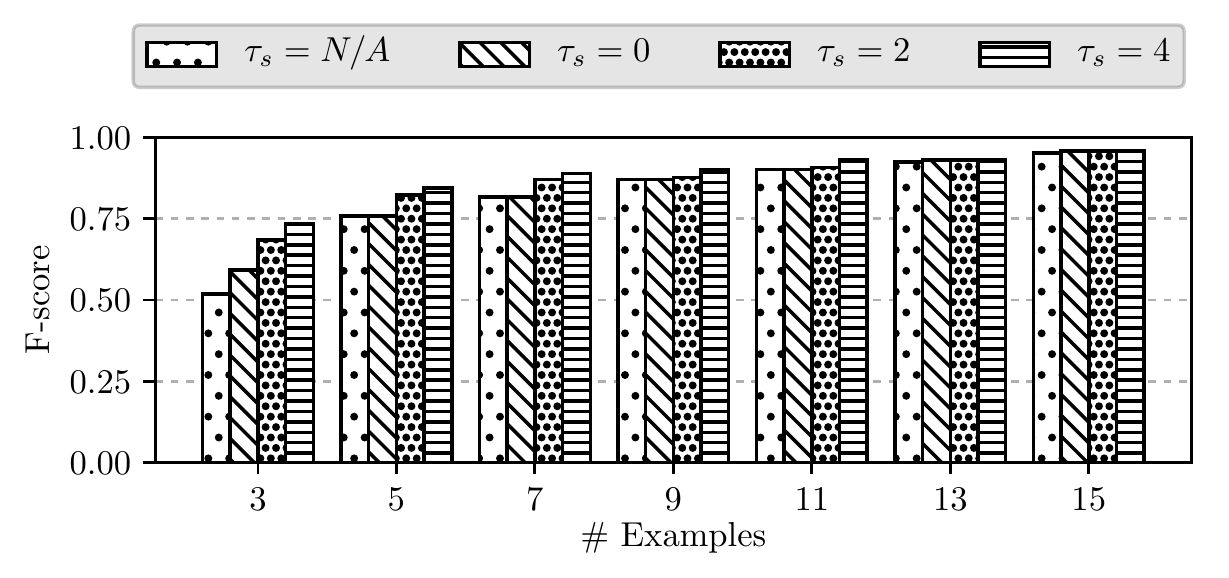}
	\vspace{-2mm}
  	\caption{Effect of different values for $\tau_s$ for \sql{IQ1} (IMDb)}
  	\vspace{2mm}
  	\label{imdbVarySKIQ1}

\end{figure}

\textbf{Scalability} expresses how the system scales when data increases. While
deciding on scalability of a system, we mark a system scalable only if it
either had a rigorous scalability experiment, or was shown to perform well on
real-world big datasets (e.g., DBpedia). Thus, we do not consider approaches as
scalable if the dataset is too small (e.g., contains 100 cells).

\textbf{Open-world} assumption states that what is not known to be true is
simply unknown. In QBE and related works, if a system assumes that tuples that
are not in the examples are not necessarily outside of user interest, follow
the open-world assumption. In contrast, closed-world assumption states that
when a tuple is not specified in the user example, it is definitely outside of
user interest.

Apart from the aforementioned metrics, we also report any \textbf{additional
requirement} of each prior art. We briefly discuss different types of
additional requirements here. \emph{User feedback} involves answering any sort
of system generated questions. It ranges from simply providing relevance
feedback (yes/no) to a system suggested tuple to answering complicated
questions such as ``if the input database is changed in a certain way, would
the output table change in this way?'' Another form of requirement involves
providing \emph{negative tuples} along with positive tuples. \emph{Provenance
input} requires the user to explain the reason why she provided each example.
Some systems require the user to provide the example tuples sorted in a
particular order (\emph{top-k}), aiming towards reverse engineering top-k
queries. \emph{Schema-knowledge} is assumed when the user is supposed to
provide provenance of examples or sample input database along with the example
tuples.

\subsection{Comparison Summary}
QBE methods on relational databases largely focus on project-join queries. Few
knowledge graph-based approaches support attribute value specification, which
is analogous to selection predicates in relational databases. However, they are
limited to predicates involving categorical attributes or simple comparison
operators ($=$ and $\neq$) involving numeric attributes. This is a serious
practical limitation as user intent is often encoded by range predicates on
numeric attributes. Therefore, we mark such limited support with the special
symbol: `!'.

While all QBE methods follow open-world assumption, QRE methods are usually
built on the closed-world assumption. However, few QRE methods also support
open-world assumption and support superset QRE variation. However, such
approaches are limited to PJ queries only. In general, QRE methods cannot
support highly expressive class of queries without severely compromising
scalability.

While almost every QBE and QRE technique supports join and projection, data
exploration techniques usually assume that the tuples reside in a denormalized
table and the entire rows of relevant entities are of user interest; thus data
exploration techniques do not focus on deriving the correct join path or
projection columns.

\subsection{Contrast with Machine Learning}\label{contrastML} 
Existing PU-learning approaches over relational data make some strong
assumptions that do not fit into our problem setting. Under the
\textbf{SCAR} assumption, TIcER~\cite{DBLP:conf/ilp/BekkerD17} estimates label
frequency, which is the sampling rate of examples, to solve the PU-learning
problem in relational data. However, when the number of positive examples is
small, it generates high-precision, but low-recall classifier. Under the
\textbf{separability} assumption, few PU-learning
approaches~\cite{DBLP:conf/kdd/YuHC02, liu2003building} infer reliable negative
examples from the positive examples and apply iterative learning to converge to
the final classifier, which is prohibitive for the real-time data exploration
setting. Aleph~\cite{srinivasan2001aleph} is a relational rule learning system
that allows a \texttt{PosOnly} setting for PU-learning, based on the
separability assumption. However, it tries to minimize the size of the
retrieved data, which results in low-recall with very few examples. Under the
\textbf{smoothness} assumption, RelOCC~\cite{DBLP:conf/aaai/KhotNS14} uses
positive examples and exploits the paths that the examples take within the
underlying relational data to learn distance measure. However, it does not
exploit any aggregated feature (deep semantic similarity) or feature statistics
(selectivity) obtained from the entire dataset. We summarize the key points to
contrast machine learning (ML) approaches with \approach below:
	 
\smallskip
	 
\noindent \textbf{Dependency on data volume}. \approach is agnostic to the
volume of unlabeled data as it relies on highly compressed summary
of the feature statistics (e.g., selectivity of the filters), precomputed over
the data. \approach pushes this summarization task in the offline
pre-processing step and uses the summary during online intent discovery. In
contrast, efficiency of ML approaches depends on the sheer volume of the data
as they are data intensive. Sampling is a way to deal with large volume of
data, however, it comes at a cost of information loss and reduced accuracy.
Moreover, for large data spread out in diverse classes, it is hard to produce
an unbiased sample; it is even harder to produce such sample in a relational
dataset. Ideally, ML approaches are task specific and the large training time
is affordable due to being a one-time requirement. However, a query intent
discovery system is designed for data exploration which demands real-time
performance. Each query intent is equivalent to a new machine-learning task and
requires time-consuming training, which is not ideal in the data
exploration setting.
	 
\smallskip
	 
\noindent \textbf{Training effort}. For each task, ML approaches require
training a new model, which requires significant effort (e.g., manual
hyper-parameter tuning) to converge to a model. Therefore, ML approaches would
need to rebuild the model every time a new query intent is posed, or even when
the current example set is augmented with new examples. No single
hyper-parameter setting would work for all tasks where the tasks are unknown
a-priori. Under the separability assumption, some PU-learning
approaches~\cite{DBLP:conf/kdd/YuHC02} apply iterative learning to converge to
the final classifier which is wasteful for learning each query intent. In
contrast, \approach does not require hyper-parameter tuning for each task,
rather it only requires one-time manual parameter tuning for the overall intent
types (e.g., user preference regarding precision-recall tradeoff) on a
particular dataset.
	 
\smallskip
	 
\noindent \textbf{Interpretability}. \approach is a query by example method
which is an instantiation of general programming by example (PBE). One key
difference between PBE and ML is the requirement of interpretability of the
underlying model. The goal of PBE technique is to provide the users the learned
model (e.g., SQL query in our case), not just a black box that separates the
intended data from the unintended one. In contrast, the focus of ML approaches
is to construct a model, often extremely complex (e.g., deep neural network),
that separates the positive data from the negative ones.
	 
\smallskip
	 
\noindent \textbf{Handling very few examples}. Even though PU-learning
approaches work with only positive examples, they require a fairly large
fraction of the positive data as examples. In contrast, \approach works on very
small set of examples which is natural for data exploration. This is possible
under the strong assumption that the underlying model, where the user examples
are sampled from, is a structured query. This implies that the user
consistently provides semantically similar examples reflecting their true
intent. When the labeled data is this small, PU-learning approaches result in
high-precision, but low-recall classifiers, which does not help in data
exploration.

\bigskip

\noindent \textbf{Assumptions involving model and feature prior}. One
significant distinction between \approach and ML approaches is the assumption
regarding the underlying model. \approach assumes that there exists a SQL query
with conjunctive selection predicates (features) that is capable to generate
the complete set of positive tuples. In contrast, ML approaches do not have any
such simplified assumption and attempts to learn a separating criteria based on
features. Hence, it is unlikely for ML systems to drop strongly correlated
features observed within the examples, despite being co-incidental.
Additionally, we exploit two information --- (1)~data dependent feature prior
(Section~\ref{contextPrior}), and (2)~data-independent feature prior (Section
~\ref{queryPriorSection}) --- which is hard to incorporate in ML. As an
example, in Section~\ref{queryPriorSection}, we discuss outlier impact, a
non-trivial component of feature prior, which basically indicates whether a set
of features together is likely to be intended. Such assumptions are hard to
encode in ML approaches.

\subsection{Contrast with Data Cube}\label{dataCube}
Data cube~\cite{dataCube} can serve as an alternative mechanism to model the
information precomputed in the abduction-ready database. However, a principal
contribution of the \adb is the determination of which information is needed
for \approach's inference, rendering it much more efficient than a data cube
solution. We have empirically evaluated data cube's performance on the IMDb
data using Microsoft SQL Server Analysis Services (SSAS) where we defined a
three-dimensional data cube: (\sql{person}, \sql{movie}, \sql{genre}), deployed
it in Microsoft Analysis Server 14 with process option ``Process Full'', and
used MDX queries to extract data. We also ported the relevant \approach \adb
data (\sql{person\-to\-genre}) from PostgreSQL into Microsoft SQL Server 14,
and evaluated the corresponding SQL queries. We found that the data cube
performs \emph{one to two orders of magnitude slower} than queries over the
\adb. One can materialize certain summary-views by applying roll-up operations
on the data cube to expedite query execution, but such materializations
essentially replicate the information materialized in the \adb; and determining
the appropriate data to materialize, i.e., \emph{which} derived relations to
precompute, is the primary contribution of the \adb.

If one were to materialize all possible roll-up operations to take advantage of
data cube's generality without the performance penalty, this would require
\emph{four orders of magnitude} more space compared to the \adb on the IMDb
data. Compression mechanisms exist to store sparse data cubes efficiently, but
such compression would hurt the query performance even more. The issue here is
that the data cube encodes non-meaningful views (e.g., \sql{person-to-movie}),
because genre is not an independent dimension with respect to movie. In
contrast, \approach aggregates out large entity dimensions (e.g., \approach
aggregates out \sql{movie} while computing \sql{person\-togenre}) which ensures
that the size of the \adb is reasonable (Figure~\ref{datasetDescription}). So,
ultimately, even though the data cube does provide a possible mechanism for
encoding the \adb data, it is not well-suited for schemas that do not exhibit
the independence of dimensions that the data cube inherently assumes, resulting
in poor performance compared to the \adb.
\end{appendix}
\fi

\end{document}